\documentclass[10pt]{article}

\usepackage{color} 
\usepackage{mathrsfs}
\usepackage{graphicx}
\usepackage{amsmath}
\usepackage{bbm}
\usepackage{amsfonts}
\usepackage{amssymb}
\usepackage{amsbsy}
\usepackage[T1]{fontenc}
\usepackage{theorem}
\usepackage{stmaryrd}
\usepackage{tipa}
\usepackage{textcomp}
\usepackage{subfigure}
\usepackage{epsfig}
\usepackage[sort,colon]{natbib}
\usepackage{lmodern}
\usepackage{framed}
\usepackage[subfigure]{tocloft}
\usepackage{subeqnarray}
\usepackage{alphalph}
\usepackage[refpage,cfg,prefix]{nomencl}
\usepackage{multirow}
\usepackage{xifthen}
\usepackage{enumerate}
\usepackage{color}
\usepackage{wasysym}
\usepackage{bibentry}
\usepackage{todonotes}
\nobibliography*
\usepackage{verbatim}
\usepackage{booktabs}
\usepackage[hyperindex,breaklinks]{hyperref}
\usepackage{array}
\hypersetup{
colorlinks=true,
urlcolor=blue,
linkcolor=black,
citecolor=black,
}
\usepackage{breakurl}
\usepackage{url}
\usepackage{setspace}
\usepackage{xr}
\usepackage{dsfont}
\usepackage{makecell}
\makenomenclature

\usepackage[labelfont=bf,labelsep=period,justification=raggedright]{caption}

\makeatletter
\renewcommand{\@biblabel}[1]{\quad#1.}
\makeatother

\date{}

\newcommand{\ud}{\mbox{d}}

\newcommand{\bs}[1]{\boldsymbol{#1}}

\newcommand{\partder}[2]{\frac{\partial #1}{\partial #2}}
\newcommand{\set}[2]{\left\{#1:#2\right\}}
\newcommand{\smallsub}[2]{#1_{\text{{\tiny #2}}}}

\newcommand{\Deltat}{\Delta t}

\newcommand{\listofalgorithms}{\textbf{\Huge{List of Algorithms}}}
\newlistof{Algorithm}{exp}{\listofalgorithms}
\newcounter{instructioncounter}
 
\newenvironment{Algorithm}[2][]
{\refstepcounter{Algorithm}
\begin{framed}\addcontentsline{exp}{Algorithm}{\protect\numberline{\theAlgorithm} #1}\par\begin{center}\textbf{Algorithm \theAlgorithm : #2}\end{center}\begin{list}
{\bf{(\arabic{Algorithm}\alph{instructioncounter}})}{\usecounter{instructioncounter}}}{\end{list}\end{framed}}
\newcolumntype{C}[1]{>{\centering\arraybackslash}m{#1}}

\begin{document}

\begin{flushleft}
{\Large
\textbf{Spatially-extended hybrid methods: a review}
}
\\
Cameron A. Smith$^{1,\ast}$, 
Christian A. Yates$^{1,\ast\ast}$
\\
\bf{1} Centre for Mathematical Biology, Department of Mathematical Sciences, University of Bath, Claverton Down, Bath, BA2 7AY, United Kingdom \\
$\ast$ E-mail: c.smith3@bath.ac.uk\\
$\ast\ast$ E-mail: c.yates@bath.ac.uk
\end{flushleft}

Key index words: hybrid modelling, reaction-diffusion, multiscale,  modelling

\section*{Abstract} \label{sect:abstract}
Many biological and physical systems exhibit behaviour at multiple spatial, temporal or population scales. Multiscale processes provide challenges when they are to be simulated using numerical techniques. While coarser methods such as partial differential equations are typically fast to simulate, they lack the individual-level detail that may be required in regions of low concentration or small spatial scale. However, to simulate at such an individual-level throughout a domain and in regions where concentrations are high can be computationally expensive. Spatially-coupled hybrid methods provide a bridge, allowing for multiple representations of the same species in one spatial domain by partitioning space into distinct modelling subdomains. Over the past twenty years, such hybrid methods have risen to prominence, leading to what is now a very active research area across multiple disciplines including chemistry, physics and mathematics. 

There  are three main motivations for undertaking this review. Firstly, we have collated a large number of spatially-extended hybrid methods and presented them in a single coherent document, while comparing and contrasting them, so that anyone with a need for a multi-scale hybrid method will be able to find the most appropriate one for their need. Secondly, we have provided canonical examples with algorithms and accompanying code, serving to demonstrate how these types of methods work in practice. Finally, we have presented papers that employ these methods on real biological and physical problems, demonstrating their utility. We also consider some open research questions in the area of hybrid method development and the future directions for the field.

\section{Introduction} \label{sect:Introduction}

The requirement for multi-scale models arises naturally from many biological and physical scenarios due to their inherent complexity. However, modelling such systems is often difficult using a single modelling paradigm. This is due to the fine balance between acquiring results in a timely manner (efficiency) and obtaining results that are consistent with the experimentally derived knowledge or physical laws (accuracy).  One such example is modelling the release of calcium from the endoplasmic reticulum, and its subsequent movement throughout the cell \citep{dobramysl2015pbd,flegg2013dsn}. Calcium ions leave the endoplasmic reticulum through ion channels which open or close depending on whether other calcium ions have bound to receptors. The behaviour of calcium ions close to the receptors can only be simulated using an individual-based method, as we require the knowledge of every particles' locations. However, when the channel opens, a large number of particles enter the cytoplasm of the cell. Keeping track of all of these particles is computationally costly, leading to limitations on the time-scales which can feasibly be simulated using the fine-grained model alone.

This review will focus on four modelling scales. The first of these is the macroscopic scale. This encompasses all models in which we make the assumption of large copy numbers within the system, such as partial differential equations (PDEs) or stochastic partial differential equations (SPDEs). In most cases, these continuum models can be simulated extremely efficiently, but they are generally invalid for low numbers of particles.

At the next finest scale is the mesoscopic scale. Typically, models at this scale employ stochastic methods in which particles are compartmentalised into small subregions of the domain, within which they are assumed to be well-mixed. Particles can transfer between compartments, and interact with other particles within their own compartment, according to a Markov chain. Models at the mesoscale can be fast to simulate with small copy numbers, but when these become large, the method can become prohibitively slow.

On an even finer scale, we have microscopic models. These simulate the trajectory of each particle in the system (typically using a fixed time-step algorithm), requiring their locations to be updated at each time-step. Examples of individual-based microsopic models include Brownian dynamics \citep{smoluchowski1917vem,andrews2004ssc} or Langevin dynamics \citep{langevin1908tmb}. These methods can be very computationally intensive. For example, for a system of $N$ particles undergoing Brownian dynamics, at each time-step, we are required to generate $\delta N$ Gaussian random variables (where $\delta$ is the dimension of the system) in order to update the positions of the particles. In addition, if pairwise interactions are necessary, the calculation of $N^2$ pairwise distances is required. For large $N$ this can be the limiting step in the method. While costly, microscopic individual-based dynamics do allow for a high level of modelling accuracy, which is often required. 

On the very finest scale are molecular dynamics \citep{holley1971mhp,durr1981mmb}. In a typical molecular dynamics simulation, a large number or particles ($\sim 10^{10}$) with attributes of mass, momentum and volume-exclusion are simulated with an extremely small time-step (typically around $10^{-15}\ s$). The position and velocity of all particles are updated according to deterministic equations specified by conservation of mass, momentum and energy. Because of the very small time-scales and enormous number of molecules, these simulations are extremely computationally expensive. However, they are necessary in order to accurately resolve the fine-level detail that is crucial for many sub-cellular processes including, for example, protein-protein interactions \citep{plattner2017cpp}.

The term `hybrid method' has come to mean many different things in the modelling literature. Typically, it refers to computational methods which represent phenomena using more than one modelling paradigm. Usually, the reason for multiple modelling paradigms is a significant separation in scale. This separation may be in time scales \citep{Cao2005sss,hellander2012cmm,klann2012hsg}, in species copy number \citep{anderson2005hmm, franz2013twh} or in spatial scales \citep{dobramysl2015pbd}. By coupling an expensive, but accurate `fine-scale' model to a cheaper, but less accurate, `coarse-scale' model, hybrid methods allow for the significant acceleration of simulations that would be computationally expensive if the fine-level model were used for all components of the system or inaccurate if the coarse-level model were employed ubiquitously.

There are range of hybrid methods that have been developed to model well-mixed systems \citep{bobashev2007hem, duncan2016hfs, hellander2007hmc, burrage2004msa, bentele2004gsh, hepp2014ahs, kiehl2004hsc, salis2005ahs}. These methods typically exploit a separation of time-scales in which fast reactions or abundant species are modelled using a coarse description and slow reactions or scarcer species are modelled using a more accurate finer description.

However, if the spatial extent of a system is important \label{page:examples}(when modelling pattern formation, travelling waves and chemotaxis \citep{murray2003mbs}, for example) then there are an even broader range of \textit{spatially-extended} hybrid methods which employ different modelling paradigms at different scales in order to complement the strengths and negate the weaknesses of each.
 
If individual species are present in very different concentrations throughout the domain (for example, in the context of chemotaxis, cells are present in low numbers whilst the chemical signalling molecules with which they interact are present in high copy numbers \citep{erban2004fic,dallon1997dcm,franz2011hmi,guo2008hab,xue2009mmt}),  distinct modelling paradigms can be used to represent each species in the same simulation. The particular representation will depend on the abundance of each species \citep{alarcon2003cam,anderson2005hmm, anderson1998cdm, dallon1997dcm, dormann2002mso, franz2011hmi, franz2013twh, gerlee2007ehc, landsberg1997gmf, jackson2006ldp, jeon2010olh,jeschke2008mrs, osborne2010ham, patel2001cam, ribba2004uhc, smallbone2007mcd, wylie2006hds}. Other types of spatial hybrid method partition the physical processes (for example reactions and diffusion) to be simulated according to their relative speeds, using a technique known as operator splitting \citep{hellander2012cmm,klann2012hsg}, simulating faster processes using relatively cheap methods and slower processes using more accurate but more expensive representations. 

For the purposes of this review, we will largely focus on methods in which distinct modelling paradigms are used in different regions of space in order to represent the same physical quantity. The models in these distinct regions of space are typically coupled together though an interface or overlap region. Spatially-coupled hybrid methods, of the sort we cover in this review, rely on the assumption that different regions of the spatial domain can be accurately represented using modelling paradigms at different scales \citep{yates2015pcm,flegg2012trm,flegg2015cmc,smith2017arm,erban2014fmd}. The motivation for these methods will typically be either a separation in the scale of species copy numbers in distinct regions of the domain or a requirement for a detailed model on small spatial scales. 

Widely differing species copy numbers in distinct regions of the domain allow coarse models to cheaply capture the dynamics in regions in which copy numbers are high whilst a fine model captures the details of low copy number populations with the required accuracy. Typically these methods would be used for phenomena which are multiscale in copy number, such as travelling wave problems \citep{moro2004hms,robinson2014atr}. Behind the wave we have large copy numbers meaning that a coarse description can be used. At the wave front and further ahead, however, stochastic variation will play a more important role in determining the correct dynamics. Consequently, a fine description is required in these regions.

Alternatively, even if there is no significant difference in copy numbers throughout the domain, there may be a small region of space which requires fine-level modelling locally, but which can tolerate coarser modelling further away in regions which are not sensitive to the individual dynamics. Typically, these methods are used to represent phenomena in which boundary effects are important \citep{dobramysl2015pbd}. 

We will refer to these methods (whatever the underlying motivating dynamics) as \textit{spatially-coupled hybrid methods}. Although we will largely focus on these spatially-coupled hybrid methods in this review, we will also touch upon other the hybrid methods which accelerate spatially-extended stochastic simulations where appropriate. 

% 
% Returning to the calcium ion channel example, in the immediate vicinity of the ion channels, an individual-based method should be used to allow for a detailed representation of the ion binding dynamics. Further away from the channels, where detailed knowledge of the particle position is less important, a coarser, cheaper scheme can be employed.

While a full description of each is beyond the scope of this review, we nevertheless reference numerous software packages designed to simulate systems at each of the four spatial scales described above (typically individually, but occasionally incorporating hybrid dynamics), which are summarised in Table \ref{tab:software}. For more information on any of these software packages, we refer the reader to the appropriate reference, which is given in the final column of the table.

\begin{table}
\centering
\begin{tabular}{|C{2.4cm}|C{5cm}|C{2.5cm}|C{3cm}|}
\hline 
\textbf{Software Package} & \textbf{Uses} & \textbf{Types} & \textbf{Reference} \\ 
\hline 
Copasi & \makecell{Next reaction method,\\ Hybrid methods} & \makecell{Meso,\\ Macro-meso} & \citet{hoops2006ccp} \\ 
\hline 
E-Cell & \makecell{Direct method,\\ Next reaction method,\\ $\tau$-leaping} & Meso & \citet{tomita1999ese} \\ 
\hline 
Lattice Microbes & \makecell{Direct method,\\ Next reaction method} & Meso & \citet{roberts2013lmh}\\
\hline
MCell & Spatial stochastic simulation & \makecell{Meso,\\ Micro} & \citet{stiles2001mcm}  \\ 
\hline 
Smoldyn & Spatial stochastic simulation & Meso-micro & \citet{andrews2004ssc} \\ 
\hline 
STEPS & Direct method & Meso & \citet{wils2009sms}\\ 
\hline
StochKit & \makecell{Direct method,\\ Optimised direct method,\\ $\tau$-leaping,\\ stochastic simulation algorithm} & Meso & \citet{li2008ass} \\ 
\hline 
(py)URDME & Next subvolume method & Meso & \citet{drawert2012urd}\\
\hline
\end{tabular} 
\caption{Summary of software implementations and the scales which they can be used to model. The table contains only packages that have been updated since 2013. All have been downloaded to test that the links still work. Adapted from \citet{pahle2009bss}.}
\label{tab:software}
\end{table}

In this paper, we review some of the vast array of hybrid methods present in the literature. In Section \ref{sect:Modelling}, we introduce the four most popular modelling paradigms for reaction-diffusion systems at different scales. In Sections \ref{sect:MacroMeso}, \ref{sect:MesoMicro} and \ref{sect:MacroMicro} we review the three main forms of spatially-coupled hybrid method. Each of these sections will begin with an in-depth review of an illustrative example, including pseudocode for its implementation, before we summarise other existing hybrid models of that type. Following these, in Section \ref{sect:Other}, several other types of hybrid method will be reviewed, before we conclude in Section \ref{sect:Discussion}.

\section{Modelling paradigms} \label{sect:Modelling} 
Within this section, we will describe modelling paradigms that are coupled most often in order to create hybrid methods. In Section \ref{sect:Modelling_macro} we describe a general PDE for reaction-diffusion systems with a single species. Section \ref{sect:Modelling_meso} contains an outline of compartment-based models, while in Section \ref{sect:Modelling_micro} we investigate individual-based dynamics. In Section \ref{sect:Modelling_Molecular}, we briefly introduce molecular dynamics, and finally in Section \ref{sect:Modelling_Equivalence} we indicate how each of these modelling methods can, in some sense, be demonstrated to be equivalent representations of reaction-diffusion.

\subsection{Macroscopic models} \label{sect:Modelling_macro}
Macroscopic models encompass ordinary differential equations (ODEs) and stochastic differential equations (SDEs) in a well-mixed context, and partial differential equations (PDEs) and stochastic partial differential equations (SPDEs) in a spatially-extended context. PDEs, with which we shall primarily be concerned in this review, are used to model the mean-field behaviour of particles, provided they are at a sufficiently high concentration, whilst SPDEs fulfil the same purpose but with the additional ability to incorporate stochasticity in particle numbers/concentrations. These macroscopic methods can be simulated efficiently, but can fail to correctly capture the appropriate behaviour at low copy numbers, in which the combination of stochastic fluctuations, small particle numbers and potentially non-linear reactions can cause significant discrepancies between the true individual-based dynamics and those of their continuum counterparts.

The methods discussed in this review which employ (S)PDEs are all designed to simulate reaction-diffusion systems, mostly comprising a single species. The PDE for the concentration of a single species, $c(\bs{x},t)$, at position $\bs{x}$ and time $t$ has the general form:

\begin{equation}
\partder{c}{t}(\bs{x},t) = D\nabla^2 c(\bs{x},t) + \mathcal{R}(c(\bs{x},t),\bs{x},t), \quad \bs{x}\in\mathbb{R}^\delta,\quad t\in[0,T]
\label{eqn:Reaction_Diffusion_Equation}
\end{equation}
with appropriate boundary and initial conditions. Here $D$ is the diffusion coefficient, $\mathcal{R}$ is a function representing the reactions and $\delta$ is the dimension of the space which we are modelling. These systems of PDEs are, in general, very difficult or impossible to solve analytically, especially when second- or higher-order reactions are involved making the reaction function $\mathcal{R}$ non-linear. Typically, however, they can be solved straightforwardly using numerical approximations. One popular family of numerical solution techniques, employed in many of the papers discussed in this review, are finite-difference methods\footnote{Note that finite-volume and finite-element methods may work equally well depending on the PDE.} such as the forward Euler or Crank-Nicolson methods.
\label{page:Finite_Difference}
Finite-difference methods discretise the spatial and temporal domains onto a mesh, upon which the PDE solution is approximated. The PDE (\ref{eqn:Reaction_Diffusion_Equation}) is converted into a system of difference equations which relate the solution at the next time-step to the solution at previous time-steps. Often, these systems of difference equations may be approximated to first order to form a linear system. There are many efficient techniques for solving such linear systems (see for example \citep{smith1985nsp,morton2005nsp,eymard2000fvm,brenner2004fem}), giving a fast method for obtaining a numerical solution of PDE \eqref{eqn:Reaction_Diffusion_Equation}.

In this review, in-keeping with the terminology used throughout the reviewed papers, these models will be described as ``macroscopic'' and, in the deterministic case as ``mean-field''.

\subsection{Compartment-based methods} \label{sect:Modelling_meso}
Compartment-based methods are a coarse-grained stochastic representation. The spatial domain is split into a number of compartments of size $h_c$, which are assumed to contain uniformly distributed, well-mixed particles. The system can be simulated using either a time-driven or an event-driven algorithm. In both cases, an event is defined as either a diffusive jump, in which a particle jumps from one compartment to a neighbour with rate $d=D/h_c^2$ (here $D$ is the corresponding macroscopic diffusion coefficient) or a reaction, in which particles interact within a compartment according to a specified reaction pathway. 

Time-driven algorithms assume a time-step, $\Delta t$, that is small enough so that at most one ``event'' occurs in the time interval $[t,t+\Delta t)$ \citep{erban2007pgs}. A scaled uniform random number is used to decide whether an event takes place, and if so, which event it is.

Event-driven algorithms are generically known in this context as stochastic simulation algorithms (SSAs). The most commonly used SSA is the Gillespie direct method \citep{gillespie1977ess}, an exact SSA in which each event, represented by a propensity function, has an exponentially distributed waiting time. Consequently, the minimum waiting time of all the  events is also exponentially distributed with a rate which is the sum of the rates of the individual reactions. The direct method, thus, simulates an exponential waiting time for the next reaction of \textit{any type} to occur and then the specific reaction to be implemented is chosen with probability proportional to its propensity function. This method is exact in the sense that it simulates the corresponding chemical master equation (CME) exactly. Although this basic method accurately simulates the underlying dynamics, it can be quite slow, and so other, faster methods have been formulated \citep{cao2004efs, elf2004ssb, gibson2000ees, li2006ldm, mccollum2006sdm,yates2013rrn}. Additionally if some moderate sacrifices in accuracy are acceptable, several approximate simulation algorithms are available, including $\tau$-leaping and $R$-leaping \citep{auger2006rla, gillespie2001aas}.

The spatially-extended methods described in this section will be referred to as ``compartment-based'', ``mesoscopic'' or ``stochastic'' (the latter only when coupled with a deterministic model) throughout this report.

\subsection{Individual-based modelling} \label{sect:Modelling_micro}

The next set of methods we will consider are individual-based methods. These methods are very computationally intensive for large numbers of particles because they require the storage and maintenance of the positions of potentially large numbers of particles. If second- or higher-order reactions or volume exclusion is to be represented, we need to consider pairwise interactions. The calculation of pairwise distances can also contribute significantly to the cost of these detailed algorithms.  In many biologically realistic situations, we may be modelling large numbers of objects at the atomistic scale. In the process of calcium induced calcium release, for example \citep{dobramysl2015pbd}, there could be tens of thousands of ion positions to keep track of, as well as millions of potential pairwise interactions.

One method of simulating diffusing particles on an individual level is to allow the particles to follow Brownian trajectories, such that: \begin{equation}
\label{eqn:BD}
\bs{y}_i(t+\Delta t) = \bs{y}_i(t) + \sqrt{2D\Delta t}~\bs{\xi}_i,
\end{equation}
 where $\bs{y}_i(t)$ is the position of particle $i$ at time $t$ and $\bs{\xi}\sim MVN(\bs{0},I_\delta)$ is a $\delta-$dimensional unit Gaussian random variable. Reactions can then be simulated in a number of different ways. One method, called the $\lambda$-$\rho$ model \citep{erban2009smr}, uses a reaction radius: if two eligible particles come within a certain distance of one another, $\rho$, they react with a given rate, $\lambda$, according to the appropriate reaction pathway. If this probability is unity and the reaction is certain to occur upon particles reaching the reaction radius, we have the special case of the ``Smoluchowski'' model \citep{smoluchowski1917vem}. Green's function reaction dynamics are an alternative event-driven microscopic model for simulating reaction-diffusion dynamics \citep{van2005gfr}, but since none of the hybrid methods discussed herein employ it, we shall not discuss it further. 

We will refer to these methods as ``individual-based'', ``microscopic'', ``particle-based'' or ``off-lattice'' models in what follows.

\subsection{Molecular dynamics} \label{sect:Modelling_Molecular}

At the very finest scale lies molecular dynamics \citep{holley1971mhp,durr1981mmb}. In molecular dynamics simulations, the molecules for the medium in which a particle of interest is moving (air, water etc.) are explicitly modelled rather than implicitly incorporated into the movement dynamics of the focal particle, as is the case with random position jumps of Brownian motion models, for example. For coarse molecular dynamics representations (as opposed to fully atomistic simulations), the particles of the medium can be considered to be identical hard spheres with a given radius and mass and whose velocity and hence momentum are specified initially, but change dynamically throughout the simulation. Particles interact with each other and in such a way as to conserve mass and momentum.

Although the resulting motion of the large focal particle may appear stochastic, it is in fact calculated deterministically by considering the many interactions with each of the small particles in the surrounding fluid, as well as the larger microscopic particles. Whilst this method of modelling explicitly accounts for the surrounding molecules instead of modelling them as a stochastic force (as in an individual-based method), keeping track of the large number of particles of the medium, their coordinates and their velocities, is computationally intensive.

\subsection{Connections between models at different scales} \label{sect:Modelling_Equivalence}

In order to couple models at different scales together, we first need to be satisfied that they are representations of the same phenomena.
Here we briefly detail how the different scale models described above can, in some senses, be thought to be equivalent to each other. We direct the interested reader to appropriate sources for full derivations. 

Firstly, in order to move from the mesoscale to the macroscale, we take the diffusive limit of a set of equations for the mean number of particles in each compartment, derived directly from the reaction-diffusion master equation \citep{erban2009smr}. In the case of second- and higher-order reactions, the mean equations depend on higher order moments (variance etc.). As a result, moment closure is required in order to close the system. The most common moment closure at first order is known as the mean-field moment-closure and the resulting equations are known as the mean-field equations. It should be noted that the mean-field PDEs derived in the case of second- and higher-order reactions, therefore, are not exact descriptions of the mean behaviour of the mesoscale model \citep{erban2007pgs}. To derive the corresponding macroscale model of diffusion from the microscale model, one can use the Fokker-Plank equation, which describes the evolution of the probability density of a particle moving according to a given SDE \citep{erban2007pgs}. For example, the Fokker-Planck equation corresponding to non-interacting particles undergoing simple Brownian motion is the canonical diffusion equation. The mesoscopic and microscopic representations can therefore be thought of as equivalent, in some sense, through their connection to the PDE. A rigorous derivation of the connections between the models at microscale and mesoscale is given by \citet{isaacson2008rbr}. Finally, the motion of a large focal particle buffetted by smaller particles of medium as part of a coarse molecular dynamics simulation, has been shown, in the limit that the focal particle's mass becomes large in comparison to the mass of the particles of the medium, to be equivalent to Brownian dynamics \citep{erban2014fmd}.

\section{Macroscopic-to-mesoscopic models} \label{sect:MacroMeso}

In this section, we will first introduce the broad concept, and then review specific examples of models which couple macroscopic dynamics to mesoscopic dynamics, which we will refer to as ``macro-meso'' hybrid methods. We list and describe the macro-meso hybrid methods covered in this section in Table \ref{table:macro_meso_summary}. We begin by giving an illustrative example of a macro-meso hybrid method, the pseudo-compartment method (PCM) \citep{yates2015pcm} and present pseudocode for its implementation. We then summarise several other existing macro-meso hybrid methods and present schematics (where appropriate) to aid the reader's understanding.

Macro-meso models are used when we want to simulate a region of the domain in which stochastic variation is important but in which the exact locations of every particle are not required, whilst for the remainder of the domain we have sufficiently high copy numbers to employ the associated continuum model. Typical examples to which these hybrid methods have been applied are the simulation of travelling wave phenomena \citep{moro2004hms, harrison2016hac}. Behind the wave-front, we have a large number of particles so that the continuum limit is valid, whilst in front of the wave, fluctuations can play a prominent role in the overall dynamics, including the wave speed.

\begin{table}
\centering
\begin{tabular}{|m{0.2\textwidth} || m{0.3\textwidth} | m{0.2\textwidth} |}
\hline
\textbf{Paper} & \textbf{Type} & \textbf{System modelled}\\
\hline
\citet{yates2015pcm} & Spatially-coupled,\newline  non-adaptive, non-overlap & Reaction--diffusion \\
\hline
\citet{moro2004hms} & Spatially-coupled,\newline  non-adaptive, non-overlap & Reaction--diffusion  \\
\hline
\citet{spill2015ham} & Spatially-coupled,\newline  adaptive, non-overlap & Reaction--diffusion \\
\hline
\citet{schulze2003ckm} & Spatially-coupled,\newline  adaptive, no-overlap & Epitaxial growth \\
\hline
\citet{harrison2016hac} & Spatially-coupled,\newline  adaptive, overlap & Reaction--diffusion \\
\hline
\citet{flekkoy2001cpf} & Spatially-coupled,\newline  non-adaptive, overlap & Reaction--diffusion \\
\hline
\citet{rossinelli2008ash} & Operator splitting & Reaction--diffusion \\
\hline
\citet{lo2016hcd} & Operator splitting & Reaction--diffusion \\
\hline
\citet{chiam2006hss} & Propensity-based spatial splitting & Reaction--diffusion \\
\hline
\end{tabular}
\caption{A summary of the macro-meso hybrid papers that will be covered in this section. The ``type'' column gives a brief description of the type of coupling used to join the two regimes. ``Spatially-coupled'' means that the domain is split into two distinct regions within which different paradigms are used. ``Adaptive'' refers to whether an interface is able to move, while ``overlap'' indicates if an overlap region is investigated. ``Operator splitting'' indicates where reaction and diffusion are modelled in different ways, rather than dividing space, and ``propensity-based spatial splitting'' is where the propensity functions are split based on their value. The ``system modelled'' column describes the application for which these models can be used. All of the macro-meso hybrid papers present novel methods rather than applications of pre-existing methods to real-world systems.}
\label{table:macro_meso_summary}
\end{table}

% (I) yates2015tpc <<<<<<

\subsection{Illustrative example of a macro-meso hybrid -- the pseudo-compartment method}\label{subsection:macro-meso_hybrid_example}

\begin{figure}[h!]
\centering
\includegraphics[width=0.8\textwidth]{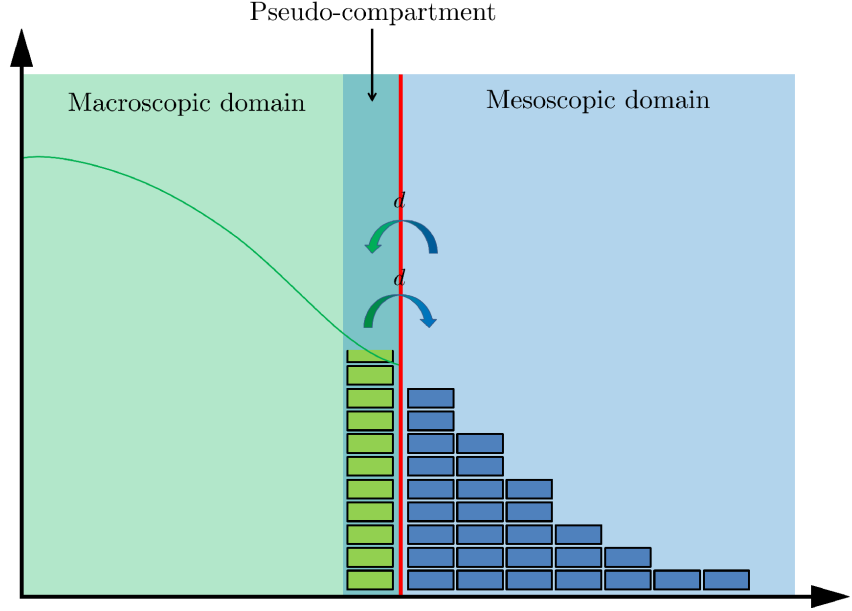}
\caption{A schematic for the PCM \citep{yates2015pcm}. The green line represents the PDE solution, while the blue boxes represent particles within each compartment. The red line denotes the interface between the two subdomains. The green boxes represent the number of pseudo-particles within the pseudo-compartment, calculated by direct integration of the solution over that region. The arrows in the centre represent the movement of pseudo-particles over the interface between the pseudo-compartment and the first compartment of the mesoscopic domain.}
\label{fig:yates2015pcm}
\end{figure}

The first macroscopic-to-mesoscopic example we present is the pseudo-compartment method (PCM) \citep{yates2015pcm}. We will treat this method as an illustrative example for this section, and as such, will present it in a high level of detail, including a schematic (see Figure \ref{fig:yates2015pcm}) and pseudo-code (see Algorithm \ref{alg:PCM}). \label{page:dimension_illustrative} Note that for all three illustrative examples, we set the dimension of space to be $\delta=1$ for simplicity.

The authors divide their domain of interest into two subdomains, separated by an interface. A PDE representation is used in one subdomain, and a compartment-based method in the other. These subdomains are labelled $\smallsub{\Omega}{P}$ and $\smallsub{\Omega}{C}$ respectively. Within the PDE subdomain, the solution is evolved using the Crank-Nicolson method (a finite-difference approximation to the underlying PDE) with zero flux boundary conditions at both ends. The time-step used for the numerical solution of the PDE is $\Delta t$ and the spatial step is $h_p$. The compartment based regime is evolved according to the Gillespie SSA, where the subdomain is split into $K$ separate compartments, each of width $h_c$, so that $|\smallsub{\Omega}{C}| = Kh_c$. The authors choose $h_c=n_ph_p$ where $n_p\in\mathbb{N}$ is the factor by which the PDE grid is finer than the compartment size. Again, a zero-flux boundary is used within $\smallsub{\Omega}{C}$ at the exterior boundary of the subdomain (i.e. the propensity for jumping out of the domain at that end is set to zero). The zero-flux boundaries on the PDE side of the interface ensure that no mass can leak from one subdomain to the other. The coupling is completed through the use of a pseudo-compartment, $C_{-1}$. This is a compartment of width $h_c$ adjacent to the interface within $\smallsub{\Omega}{P}$. A schematic for this method is shown in Figure \ref{fig:yates2015pcm}.

Pseudo-particle numbers within this pseudo-compartment are calculated through direct integration of the PDE, giving 
$$ n\left(C_{-1},t\right)=\int_{C_{-1}}{c(x,t)\ dx},$$
where $n(A,t)$ is the number of particles residing in the region $A\subseteq\Omega$ at time $t$. This value is then used to generate a propensity function for particles jumping out of the pseudo-compartment and into the first compartment adjacent to the interface in $\smallsub{\Omega}{C}$. Similarly, in order to correctly model the flux over the interface, particles in the first compartment in $\smallsub{\Omega}{C}$ can jump into the pseudo-compartment with the usual diffusive rate. 

The algorithm proceeds by firstly generating a time until the next event (a diffusive jump between (pseudo-)compartments or one of the $M$ reactions within the true compartments) according to the Gillespie algorithm \citep{gillespie1977ess}. This can be found by transforming a uniform random variable $u_1\sim\text{Unif}(0,1)$ into an exponential random variable with rate equal to the sum of all propensity functions, given by 
\begin{equation}
\tau = \frac{1}{\alpha_0}\ln\left(\frac{1}{u_1}\right),
\label{eqn:next_event_time}
\end{equation}
where $\alpha_0$ is the sum of all propensity functions (including the extra ones for jumps out of and into the pseudo-compartment). The algorithm then checks to see whether the time has been incremented past the next PDE update time. If not, a compartment-based event occurs first, and an event is selected with probability proportional to its propensity function. Otherwise, the numerical solution of the PDE is incremented by a single time-step. When a particle jumps from the pseudo-compartment to the first compartment of $\smallsub{\Omega}{C}$, we remove a particle's worth of mass uniformly from the PDE solution at the points within the pseudo-compartment, and increment the count of particles in the first compartment. A movement in the opposite direction is completed in a similar manner, by adding a particle's worth of mass to the PDE solution uniformly across the pseudo-compartment, and removing a particle from the first compartment. Pseudocode for this method is given in Algorithm \ref{alg:PCM}.

\begin{Algorithm}{Pseudo-compartment method (PCM)} \label{alg:PCM}
 \item Initialise the time, $t=t_0$ and set the final time, $T$. Specify the PDE-update time-step $\Deltat$ and initialise the next PDE time-step to be $t_\Delta=t+\Deltat$.
 \item Initialise the number of particles in each compartment in $\smallsub{\Omega}{C}$, $n(C_i,t)$ for $i=1,\dots, K$ (where $C_i$ is the region of the domain covered by compartment $i$), and the distribution of density in $\smallsub{\Omega}{P}$, $c(x,t)$, for $x\in\smallsub{\Omega}{P}$.
 \item \label{item:calculate_the_propensity_functions} Calculate the propensity functions for diffusion between the compartments as $\alpha_{i,j}=n(C_i,t)D/h_c^2$ for $i=1\dots K$ and $j=M+1, M+2$ (corresponding to left and right movements) and for reactions as $\alpha_{i,j}$ for $i=1\dots K$ and $j=1,\dots, M$ using the usual mass action kinetics.
 \item Calculate the propensity function for diffusion from the pseudo-compartment, $C_{-1}$, in $\smallsub{\Omega}{P}$, into the adjacent compartment, $C_1$, in $\smallsub{\Omega}{C}$: $\alpha^*=D\int_{C_{-1}}c(x,t)\ \ud x/h_c^2$.
 \item Calculate the sum of the propensity functions, $\alpha_0=\sum^K_{i=1}\sum^{M+2}_{j=1}\alpha_{i,j}+\alpha^*$.
 \item Determine the time for the next `compartment-based' event, $t_c=t+\tau$, where $\tau$ is given by equation \eqref{eqn:next_event_time}.
 \item \label{item:update_particle_numbers_in_compartments} If $t_c<t_\Delta$ then the next compartment-based event occurs:
 \begin{enumerate}[(a)]
 \item Determine which event occurs according to the method described in the text (see \citet{gillespie1977ess}).
 \item If the event corresponds to $\alpha_{i,j}$ for $i=1\dots K$ and $j=M+1,M+2$ then move a particle from interval $i$ in the direction specified by $j$. If the particle crosses the interface into pseudo-compartment, $C_{-1}$, then add a particle's worth of mass uniformly to the region $C_{-1}$ i.e. $c(x,t+\tau)=c(x,t)+\mathds{1}_{[x\in C_{-1}]}/h_c$. Here, $\mathds{1}_{x\in A}$ is an indicator function which takes the value 1 when $x\in A$ and 0 otherwise.
  \item \label{item:remove_mass_from_PDE} If the event corresponds to propensity function $\alpha^*$ and $c(x,t)>1/h_c$ for all  $x\in C_{-1}$ then place a particle in $C_1$. Remove a particle's worth of mass from the PDE solution in the  region $C_{-1}$ i.e. $c(x,t+\tau)=c(x,t)-\mathds{1}_{[x \in C_{-1}]}/h_c$.
  \item Update the current time, $t=t_c$.
 \end{enumerate}
\item If $t_\Delta<t_c$ the the PDE regime is updated:
\begin{enumerate}[(a)]
\item Update the PDE solution according to the numerical method. 
 \item Update the current time, $t=t_{\Delta}$ and set the time for the next PDE update step to be $t_{\Delta}=t_{\Delta}+\Deltat$.
\end{enumerate}
\item If $t\leq T$, return to step \ref{item:calculate_the_propensity_functions}.

Else end.

\end{Algorithm}

\begin{figure}[h!]
	\begin{center} 
	\subfigure[][]{
		\includegraphics[width=0.31\textwidth,trim={32pt 0pt 35pt 0pt},clip]{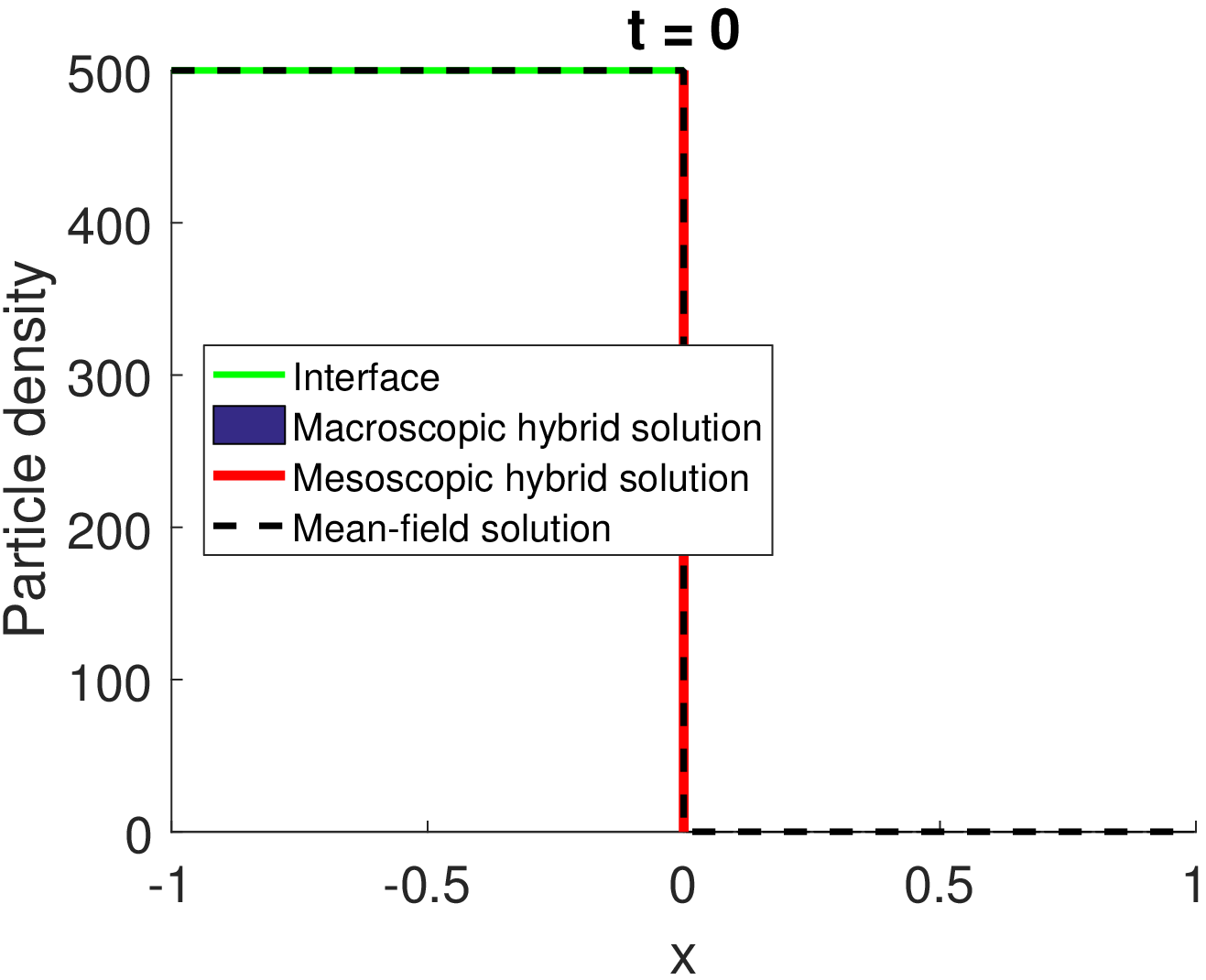}
		\label{fig:yates2015pcm_1}
	}
	\subfigure[][]{
		\includegraphics[width=0.31\textwidth,trim={32pt 0pt 35pt 0pt},clip]{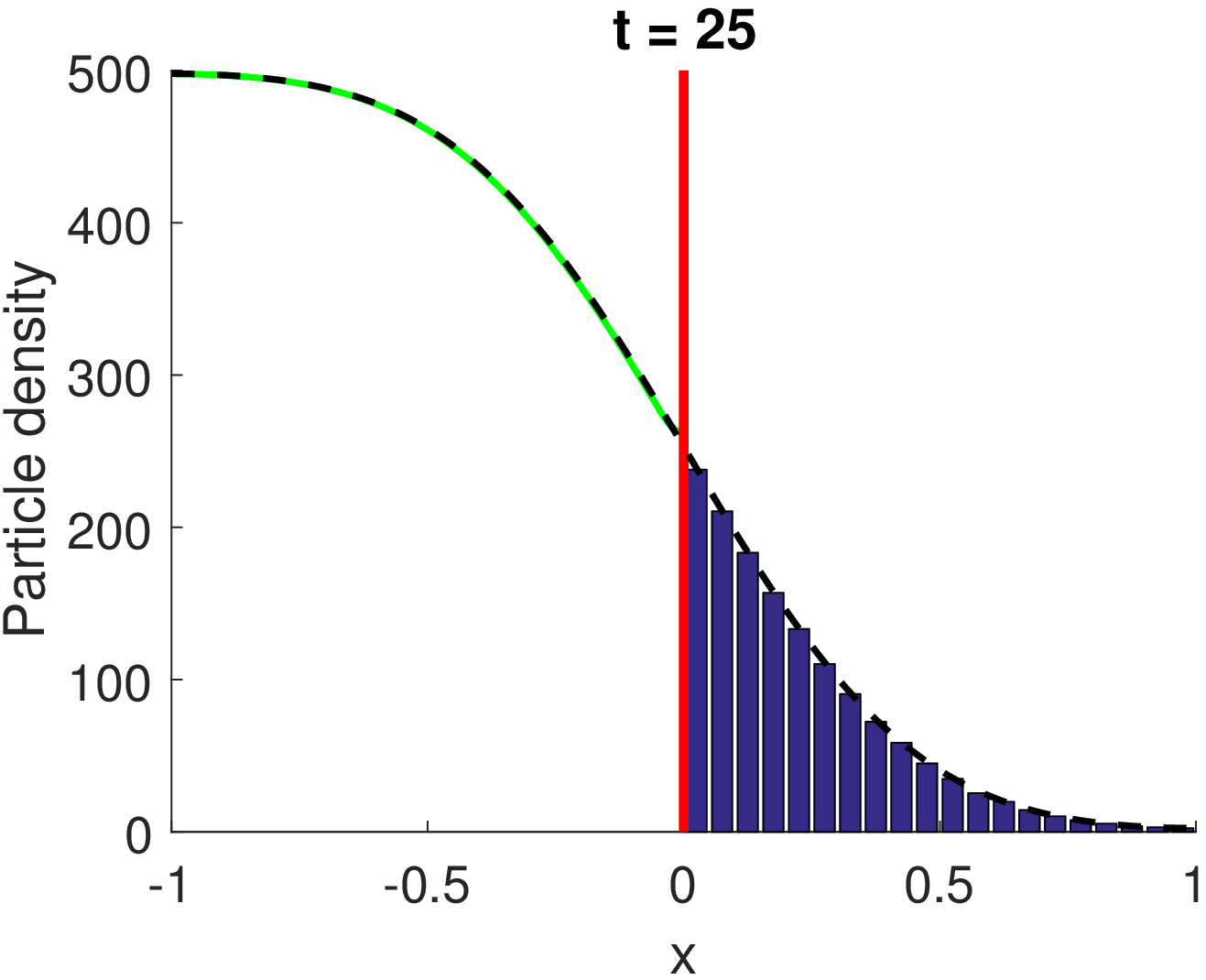}
		\label{fig:yates2015pcm_2}
	}
	\subfigure[][]{
		\includegraphics[width=0.31\textwidth,trim={32pt 0pt 35pt 0pt},clip]{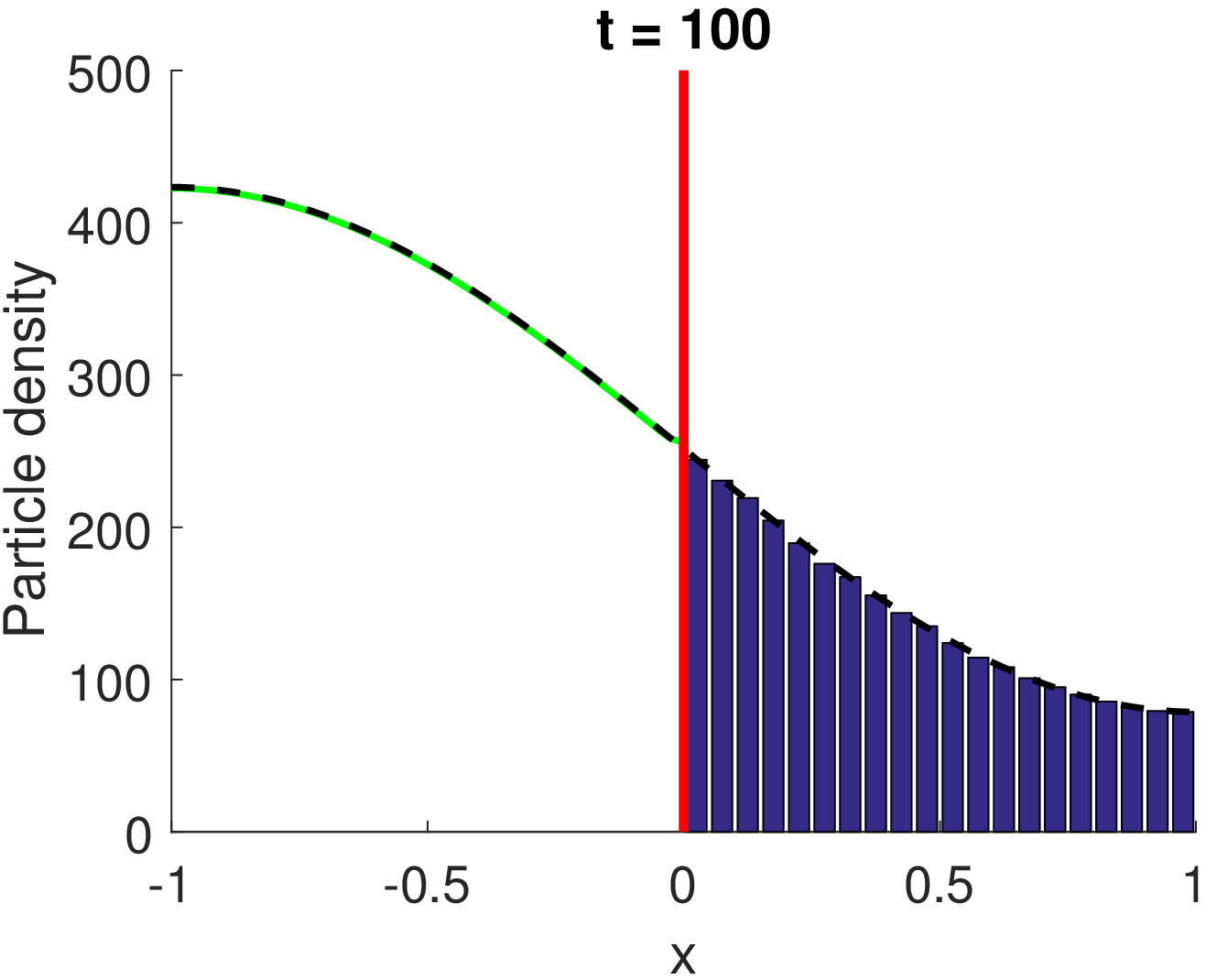}
		\label{fig:yates2015pcm_3}
	}
	\caption{A replication of results from \citet{yates2015pcm} using the PCM. The green line corresponds to the PDE part of the hybrid solution, the red line is the interface, the blue bars are the compartment-based part of the hybrid solution. The dashed black line is the analytical solution of the mean-field PDE model (the diffusion equation) across the entire domain. Parameter values are as in the text.}
	\label{fig:yates2015pcm_example}
\end{center}
\end{figure}

In Figure \ref{fig:yates2015pcm_example} we have reproduced an example simulation from \citep{yates2015pcm} using the pseudo-compartment method. We initialise $N=500$ particles uniformly throughout the PDE subdomain, where $\smallsub{\Omega}{P}=(-1,0)$ and $h_p=0.01$. The compartment-based subdomain, $\smallsub{\Omega}{C}=(0,1)$, is split into $K=20$ compartments, each of width $h_c=0.05$. The interface naturally lies at $I=0$ and the results were averaged over 5000 repeats until a final time of $T=100$. We set the diffusion coefficient to be $D = 0.0025$ and the PDE time-step to be $\Deltat = 0.01$.

% moro2004hms
\subsection{Other macro-meso hybrid methods}\label{section:other_macro-meso_hybrid_models}

We now turn our attention to other macro-meso hybrid methods, indicating where they share similarities with one another and where they differ. The full list of methods considered in this section is given in Table \ref{table:macro_meso_summary}.

Another type of hybrid method incorporates an adaptive interface. The interface between two modelling regions moves adaptively based on a pre-determined criteria, that may involve (local) copy numbers or densities. \citet{moro2004hms} present one such hybrid method when investigating pulled fronts in a diffusive reversible dimerisation. In contrast to the PCM above, they use the same discretisation for both the continuum and the compartment-based simulations. The boundary between the two subdomains is determined using a threshold number of particles. Any voxels with more particles than this threshold will be simulated by numerically solving the macroscopic Fisher-Kolmogorov-Petrovsky-Piscounov (FKPP) equation. Any voxels with fewer than this number of particles are simulated as a mesoscopic compartment-based position-jump Markov chain. If particles in the compartment-based region jump into the macroscopic region, they are immediately removed from their voxel and held until the next PDE update step. When the PDE update occurs, PDE voxels away from the interface are updated according to the usual finite-difference method, but the value of the voxel closest to the interface is updated with a mixed flux condition. Flux from the macroscopic side to the mesoscopic side is specified by the deterministic flux from the PDE region, whereas flux from the mesoscopic side to the macroscopic side is determined by the number of particles that jumped beyond the interface into the macroscopic subdomain from the mesoscopic subdomain during the PDE update time-step. Flux in the opposite direction (from macroscopic to mesoscopic) is implemented by adding a Poisson distributed random number of particles (with mean corresponding to the expected flux of particles over the boundary as determined by the deterministic model) to the first voxel in the mesoscopic region.

% (I) spill2015ham

\begin{figure}[h!]
\centering
\includegraphics[width=0.8\textwidth]{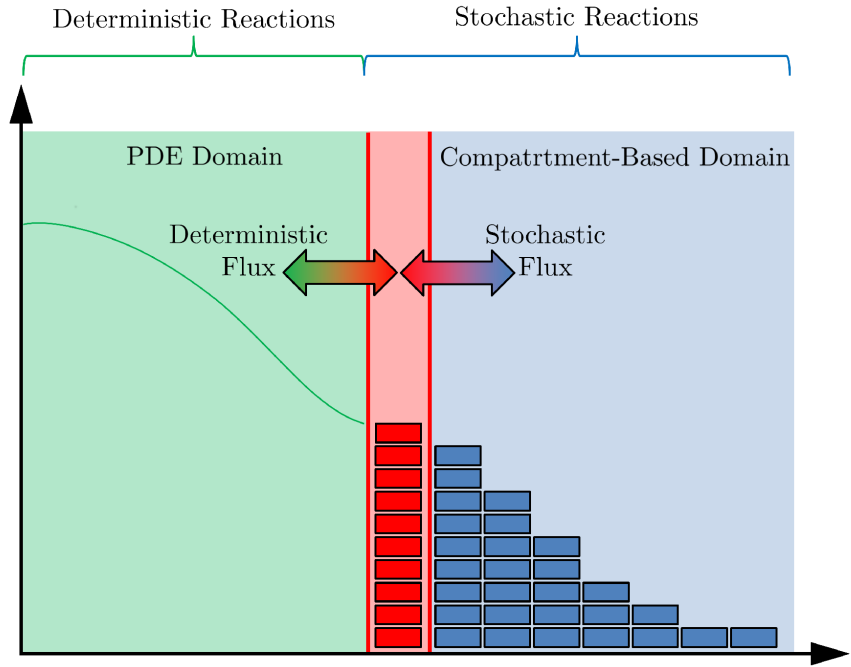}
\caption{A schematic for the method from \citet{spill2015ham}. The green line and blue boxes are as in Figure \ref{fig:yates2015pcm}, while the red boxes denote an extra compartment between the PDE and compartment subdomains. The coloured double-headed arrows denote how the flux over each of the two red interfaces are calculated.}
\label{fig:spill2015ham}
\end{figure}

Building upon this idea of adaptive interfaces, \citet{spill2015ham} include the possibility of having multiple adaptive interfaces (see Figure \ref{fig:spill2015ham} for a schematic with a single interface). As in \citet{moro2004hms}, the same grid spacing is used for both modelling paradigms. The authors are able to add multiple interfaces by again introducing a threshold value in order to determine which regions of the domain should be simulated deterministically and which stochastically, allowing the positions of the interfaces between distinct modelling regions to move, appear and disappear. Boxes with particle numbers lower than the threshold are simulated according to the compartment-based dynamics. Boxes with particle numbers greater than the threshold are categorised as deterministic and evolve according to a set of coupled ODEs which describe the mean field number of particles in each compartment. The single threshold value potentially gives rise to multiple distinct regions of stochastic and deterministic modelling for species whose values fluctuate around the threshold value. In order to ensure there are not too many distinct regions a minimum subdomain size condition is implemented which prevents the occurrence of small, disconnected regions of a particular method. 

To implement the coupling between the macroscale and mesoscale models, flux from the deterministic side is governed by the mean-field ODEs, while particles can jump into and out of the interface compartment from the mesoscopic side with rates determined by the SSA \citep{gillespie1977ess} (in a method similar to that of the PCM \citep{yates2015pcm}). All reactions within the interface compartment are completed using the SSA, whereas reactions in other parts of the domain are implemented according to their respective modelling paradigm.

% schulze2003ckm

Although many hybrid methods are designed for simulating reaction-diffusion systems, others have been designed to represent different physical phenomena. \citet{schulze2003ckm} present a hybrid method for modelling epitaxial growth. The method couples a discretised version of the macroscopic Burton-Cabrera-Frank (BCF) continuum model for the growth of a crystalline structure to its corresponding, on-lattice, mesoscopic kinetic Monte-Carlo (KMC) representation. In this mesoscopic model, crystals grow layer upon layer. Layers are first nucleated and then expand by the addition, surface diffusion, and deposition of adatoms (crystalline particles) from solution. The front of a growing layer is referred to as a ``step''. The method for simulating the KMC model is taken from \citep{bortz1975nam}, however, it proceeds in the same way as the Gillespie SSA \citep{gillespie1977ess}. The BCF model, as implemented in this paper, is effectively a finite-difference discretisation of the diffusion equation. This continuum representation is employed in cells which comprise multiple sites of the individual-based model. Steps are simulated using the fine-grained KMC algorithm, and regions away from steps are simulated using the coarse diffusion approximation for the movement of adatoms on the surface. Separating the subdomains are interfaces, which adaptively move with the locations of the steps. The authors consider both two- and three-dimensional simulation regions, referred to as the 1+1- and 2+1-dimensional domains (the ``$+1$'' refers to the crystals growing upwards, meaning that we are effectively simulating a surface process in one- and two-dimensional space).

The algorithm proceeds in a similar way to the PCM \citep{yates2015pcm} for reaction-diffusion systems. Close to a step, adatoms are represented using the stochastic KMC algorithm so that their locations can be individually updated, and processes such as absorption, dissociation and nucleation can be accurately modelled. Further away from a step, we neglect these processes and simply consider the particles diffusing along the surface. The time until the next KMC event is calculated using exponentially distributed random variables. If the next KMC event occurs before the next PDE update time, the corresponding event is enacted, otherwise the PDE is evolved forwards in time. Particles jump across the interface, with a rate which depends on the number of particles within the continuum cell adjacent to the interface. These stochastic jump events are simply added to the list of KMC events. If a particle leaves the continuum cell, a new particle is initialised in an adjacent KMC site and the density in the continuum cell is decreased uniformly across its width by a total of one particle. In the opposite direction, the particle is removed from the KMC simulation and a particle's worth of mass is added uniformly across the corresponding continuum cell. As with the PCM, care has to be taken to ensure positive density in the continuum at all times. The interface is also adaptive in that it can evolve as the steps move through space. If a cell needs to change representation from KMC to BCF, we simply count the number of particles in this region and convert it to a particle density uniformly spread across the now-continuum cell. In the opposite direction, the density is converted to the floor of the number of particles (whilst remembering the fractional part in case the cell is again represented by the continuum description later in the simulation). This number of particles is then initialised randomly throughout the now-discretised cell.

% (I) harrison2016hac

\begin{figure}[h!]
\centering
\includegraphics[width=0.8\textwidth]{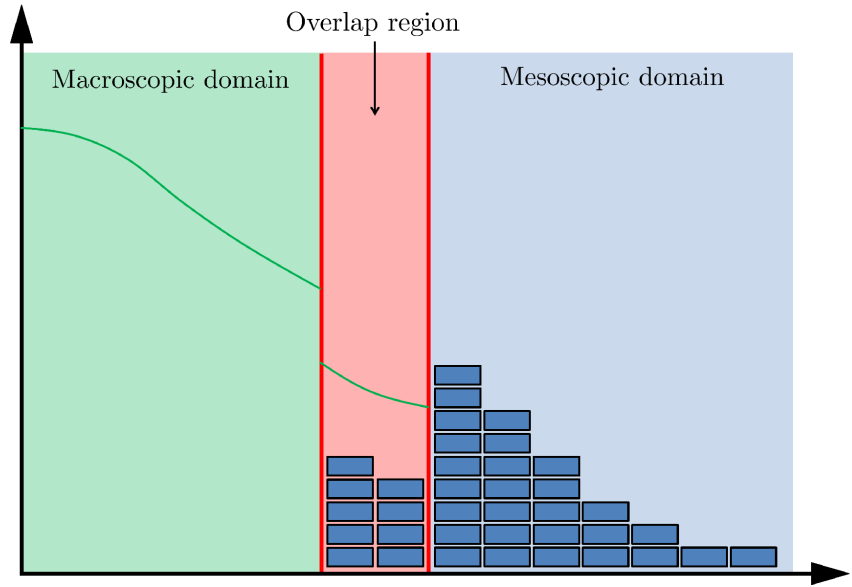}
\caption{A schematic for the method of \citet{harrison2016hac}. The descriptions for the green line and blue bars are the same as in Figure \ref{fig:yates2015pcm}. The overlap region is denoted by the red region. The width of the overlap region can be any integer number of compartment widths (here, for simplicity, we have chosen a two compartment-width overlap region). In the overlap region, the sum of the densities of the two methods gives the overall solution.}
\label{fig:harrison2016hac}
\end{figure}

Point interfaces are not the only way to divide the domain between modelling paradigms -- overlap regions may also be employed. Typically these regions inherit properties from both of the models that are being coupled. \citet{harrison2016hac} utilise such a region to couple their mesoscopic and macroscopic models of reaction diffusion. The authors suggest a fixed-time-step, finite-difference scheme for the numerical solution of the macroscopic PDE and use a time-driven algorithm for simulating the stochastic regime (with the same fixed time-step as the PDE). This is in contrast to many of the other hybrid algorithms within this review, in which the Gillespie SSA \citep{gillespie1977ess} is employed for the mesoscopic regime. It is noted, however, that event-driven alternatives can be applied with minor alterations. 

The authors focus on reaction-diffusion systems in one dimension with the compartment-based subdomain on the right and the PDE subdomain on the left (see Figure \ref{fig:harrison2016hac}) (although the algorithm would work equally well in higher dimensions and with the orientation of the regions reversed). The overlap region has two interfaces, one at either end. At the right-hand interface where the PDE begins (part-way into the compartment subdomain), a Dirichlet matching boundary condition is implemented on the PDE. This is achieved by calculating the average concentration in the two compartments either side of the interface, and ensuring that the PDE solution at the interface is set to that value. At the left-hand interface, where the compartment-based subdomain ends (part-way into the PDE subdomain), a flux-matching boundary condition is applied to the compartment immediately to the right of the interface. The diffusive flux across the interface is calculated using the value of the PDE lattice sites corresponding to the centres of compartments either side of the interface. This flux is then imposed on the compartment-based regime by adding or removing particles from the left-most compartment with probability proportional to the magnitude of the flux (with time-step chosen to ensure this magnitude is less than one). An adaptive interface condition similar to that implemented in the adaptive two-regime method \citep{robinson2014atr} (see Section \ref{section:other_meso-micro_hybrid_models}) is also presented. Repositioning criteria based on density are checked at pre-defined time-steps, and the overlap region is moved accordingly.

% flekkoy2001cpf

Similarly to \citet{harrison2016hac}, \citet{flekkoy2001cpf} utilise an overlap region as part of a non-adaptive algorithm. They introduce a method for coupling a discretised version of the diffusion equation with a discrete-time and -space mesoscopic Markov chain representation of diffusion in which particles can jump to neighbouring voxels in each fixed time-step. The PDE time-step is chosen to be coarser than its stochastic counterpart, meaning that there can be multiple stochastic jumps for every PDE update step. The spatial-mesh for the mesoscopic, stochastic representation is also finer than that of the corresponding discretisation of the diffusion equation; that is to say that there are multiple mesoscopic voxels for every macroscopic voxel. This is in contrast to many of the other macroscopic-to-mesoscopic coupling methods we have outlined in this review, in which the PDE mesh is at least as fine as the compartment size. In these papers, this finer macroscopic resolution was motivated by the idea that the PDE is an exact representation of the scaled probability density of diffusing particles and so warranted an appropriately fine discretisation. Here, \citet{flekkoy2001cpf} motivate their choice of discretisation (multiple mesoscopic voxels for every macroscopic voxel) by arguing that the PDE-based model is a coarse-grained version of the particle model and hence requires a coarser discretisation in both space and time.

In order to couple the two methods, \citet{flekkoy2001cpf} allow the two subdomains to overlap across several PDE sites. Within this overlap region, mass is represented as both mesoscopic and macroscopic. The regimes are coupled using a flux-balancing argument which implements the flux of the macroscopic representation on the mesoscopic model at one end of the overlap region and vice versa at the other. The flux term from the PDE description is implemented as a source term which is added to the particle description on the penultimate mesoscopic mesh point. This PDE flux is calculated by using a centred finite-difference approximation across the two PDE sites which span the penultimate mesoscopic mesh point. However, in order to prevent discontinuities in density between the different descriptions, the \textit{PDE} density at one of the two mesh points (used in the finite-difference approximation of the PDE gradient) is substituted for the \textit{particle} density at the same point. At the other end of the overlap region, the averaged particle flux (determined to be the difference between the number of right moving and left moving particles) over a PDE time-step is added to the penultimate site of the PDE mesh.

% rossinelli2008ash

The previous six methods detailed in the macro-meso section \citep{yates2015pcm,moro2004hms,spill2015ham, schulze2003ckm,harrison2016hac,flekkoy2001cpf}\label{page:six_methods} are all spatially-coupled hybrid methods -- methods that split the spatial domain into distinct (possibly partially overlapping) regions in which different modelling methods are used. However, other methods exist, which do not specify distinct or even overlapping subdomains for each of the two methods to be coupled. We now focus on two other types of hybrid method. The first employs operator splitting - a process in which the operators which evolve the system are implemented separately \citep{rossinelli2008ash,lo2016hcd}. The second method employs propensity-based spatial splitting \citep{chiam2006hss}, which divides the representation of the dynamics adaptively according to the value of each event's propensity function.

\citet{rossinelli2008ash} use $\tau$-leaping \citep{gillespie2001aas} in order to introduce two new methods for accelerating stochastic reaction-diffusion systems \citep{cao2006ess}. The spatial domain is discretised into a regular lattice, with the particles situated at each lattice site subject to the same reactions. Particles can also diffuse to neighbouring lattice sites with appropriately chosen rates.

The first accelerated method presented by \citet{rossinelli2008ash} is a purely stochastic algorithm that the authors name the ``spatial $\tau$-leap'' ($S\tau$-leap) method. This is not a hybrid method, but does allow for faster approximate simulations by employing $\tau$-leaping. This algorithm proceeds by calculating maximum acceptable leap times for reactions and diffusive events across all voxels. The minimum of these adaptively chosen, acceptable times, $\tau$, is then selected as the next time-step for the algorithm. The entire system is updated by drawing Poisson random variables to simulate the number of events of each type that occur during the next $\tau$ time units.

The second method \citet{rossinelli2008ash} introduce is the ``hybrid $\tau$-leap'' ($H \tau$-leap) method. This method exploits the premise that diffusion processes are typically up to two orders of magnitude faster than corresponding reaction processes \citep{bernstein2005smr}. For this method, the authors split the dynamics, completing the diffusive jumps deterministically and the reactions using the $\tau$-leaping method. The time-step for the reactions is calculated adaptively, as before, but only the reactions are updated in this step. Following this, a centred finite-difference approximation combined with forward Euler time-integration is used to deterministically advance the diffusion of particles according to the macroscopic diffusion operator.

% lo2016hcd

A similar operator splitting method is presented by \citet{lo2016hcd}. Their method simulates all reactions using a compartment-based mesoscopic representation, implemented using the Gillespie SSA \citep{gillespie1977ess}. Where molecule numbers are sufficiently large, the number of diffusive jumps between compartments are approximated using continuous Gaussian random variables, with time-dependent means and variances. Where particle numbers are low, diffusive jumps are implemented as events within the SSA. This coupling allows for large time-steps to be taken, even in the presence of rapid diffusion. The numbers of diffusive jumps between compartments are approximated as the sum of the ``deterministic'' number of jumps and appropriately scaled zero-mean Gaussian random variables. The system size expansion is applied to the reaction-diffusion master equation (RDME) in order to characterise the covariances of these random variables. 

%The algorithm proceeds as follows. Firstly, the time until the next reaction that occurs is calculated using a standard SSA technique. The ``now-continuous'' number of diffusion events between compartments is then calculated over this time-frame (assuming that no reaction events occur) using the Gaussian approximation. If the expected number of diffusion events for a particular compartment is above a pre-determined threshold then the Gaussian approximation is applied to calculate the number of particles leaving that compartment. Otherwise diffusion is implemented using the SSA. In this way, the compartments in which a Gaussian approximation is used may change adaptively as the simulation progresses.
% chiam2006hss

Another type of hybrid method chooses which events of the compartment-based regime are to be simulated using the continuum or mesoscopic solvers by using their propensity functions. \citet{chiam2006hss} simulate the mesoscopic dynamics using the Gillespie SSA \citep{gillespie1977ess} while the PDE is discretised using a second-order finite-difference approximation and evolved using the forward Euler method. Each of these descriptions is simulated on the same discretised mesh. Propensity functions are calculated for all possible events (reactions within and diffusive jumps from each box). A threshold value is then used to decide which events are to be simulated using the SSA and which using the deterministic description. The threshold value corresponds to a given fraction of the maximum propensity function. Any events with a sub-threshold propensity are simulated using the SSA. Those with super-threshold propensities are simulated using the finite-difference discretisation. The authors comment that the value of the threshold needs to be ``tuned'' depending on the specific problem to obtain the correct balance between efficiency and accuracy.  

% \textcolor{red}{There are several sources of error associated with the algorithm. The most prominent of which is related to the implementation of diffusion events. Even for a single species, differences in copy number across the domain may lead to propensity functions which suggest the classification of diffusive jump events in one part of the domain to be deterministic and in another to be stochastic. In this case diffusion is implemented across the whole domain using the finite-difference discretisation as well as using the compartment-based method for the compartments in which this species' population is sufficiently low ***I think this sentence is a bit confusing. Can we revisit it? Why is everything moved according to the deterministic method and then moved again according to the stochastic method. Might be best to cut this paragraph if it is too confusing.***. This has the potential to lead to excessive diffusion of this species DONT THINK THIS IS NEEDED, ALL INFO IS IN TOP PARAGRAPH. NOT REALLY SURE WHAT THIS PART IS TRYING TO SAY.}

In this section we have outlined several spatially-extended hybrid methods which can be used to couple macroscopic and mesoscopic methods. We now turn our attention towards mesoscopic-to-microscopic couplings.

\section{Mesoscopic-to-microscopic models} \label{sect:MesoMicro}

In this section we will begin by introducing, in broad terms, models which couple microscopic dynamics to mesoscopic dynamics, which we will refer to as ``meso-micro'' hybrid methods. After summarising the key properties of the meso-micro hybrid methods covered in this section, in Table \ref{table:micro_meso_summary}, we go on to describe them in more detail. We begin by giving  a detailed description of an illustrative example of a meso-micro hybrid method, the ghost cell method \citep{flegg2015cmc} and present pseudocode for its implementation. We then summarise other existing meso-micro hybrid methods.

% kalzantis2009hss (rubbish one) Opening paragraph
For meso-micro hybrid methods, both of the models which comprise the hybrid method incorporate some form of stochastic variation. These types of method will be required whenever fluctuations are deemed important across the entire domain, but where specific particle locations are not required in some subregions of the domain. As an example, we can consider the modelling of an ion channel \citep{dobramysl2015pbd,flegg2013dsn}. We require detailed knowledge of the molecules in regions of space close to the ion channel's receptors in order to resolve the binding dynamics accurately. However, away from the channels, this detailed representation is not required.

\begin{table}[h!]
\centering
\begin{tabular}{|m{0.2\textwidth} || m{0.3\textwidth} | m{0.2\textwidth} |}
\hline
\textbf{Paper} & \textbf{Type} & \textbf{System modelled}\\
\hline
\citet{flegg2015cmc} & Spatially-coupled,\newline  non-adaptive, non-overlap & Reaction--diffusion \\
\hline
\citet{flegg2012trm} & Spatially-coupled,\newline  non-adaptive, non-overlap & Reaction--diffusion \\
\hline
\citet{robinson2014atr} & Spatially-coupled,\newline adaptive,
 non-overlap & Reaction--diffusion \\
\hline
\citet{flegg2014atr} & Spatially-coupled,\newline  non-adaptive, non-overlap & Reaction--diffusion \\
\hline
\citet{dobramysl2015pbd} & Spatially-coupled,\newline  non-adaptive, non-overlap & Reaction--diffusion\\
\hline
\citet{hellander2012cmm} & Operator splitting & Reaction--diffusion \\
\hline
\citet{klann2012hsg} & Operator splitting & Reaction--diffusion \\
\hline
\end{tabular}
\caption{A summary of the meso-micro hybrid papers that will be covered in this section. The methods in all the meso-micro hybrid papers summarised here are designed for modelling reaction-diffusion systems. Each of these papers are concerned with the development of a novel hybrd method, apart from the paper by \citet{dobramysl2015pbd}, which employs the two-regime method \citep{flegg2012trm} to investigate the formation of calcium puffs. See text for more information. Descriptors are as in Table \ref{table:macro_meso_summary}.}
\label{table:micro_meso_summary}
\end{table}

% (I) flegg2015cmm
\subsection{Illustrative example of a meso-micro hybrid -- the ghost cell method}

%\citet{flegg2015cmc} compare the TRM with a method that they have developed, called the ghost cell method. This method is also an interface based method, but all jumps over the interface are governed by the compartment-based scheme. The authors accomplish this by creating a ``ghost-cell'' adjacent to the interface but within the microscopic subdomain, which acts as a compartment. The particles that fall in this region are counted to give the number of particles within the ghost-cell, and particles can jump between this and the final compartment of the mesoscopic subdomain using the SSA approach. This means that no problems can arise from moving a particle from the microscopic subdomain into the mesoscopic subdomain and vice versa because the flux is governed purely by one representation. 

\begin{figure}[h!]
\centering
\includegraphics[width=0.8\textwidth]{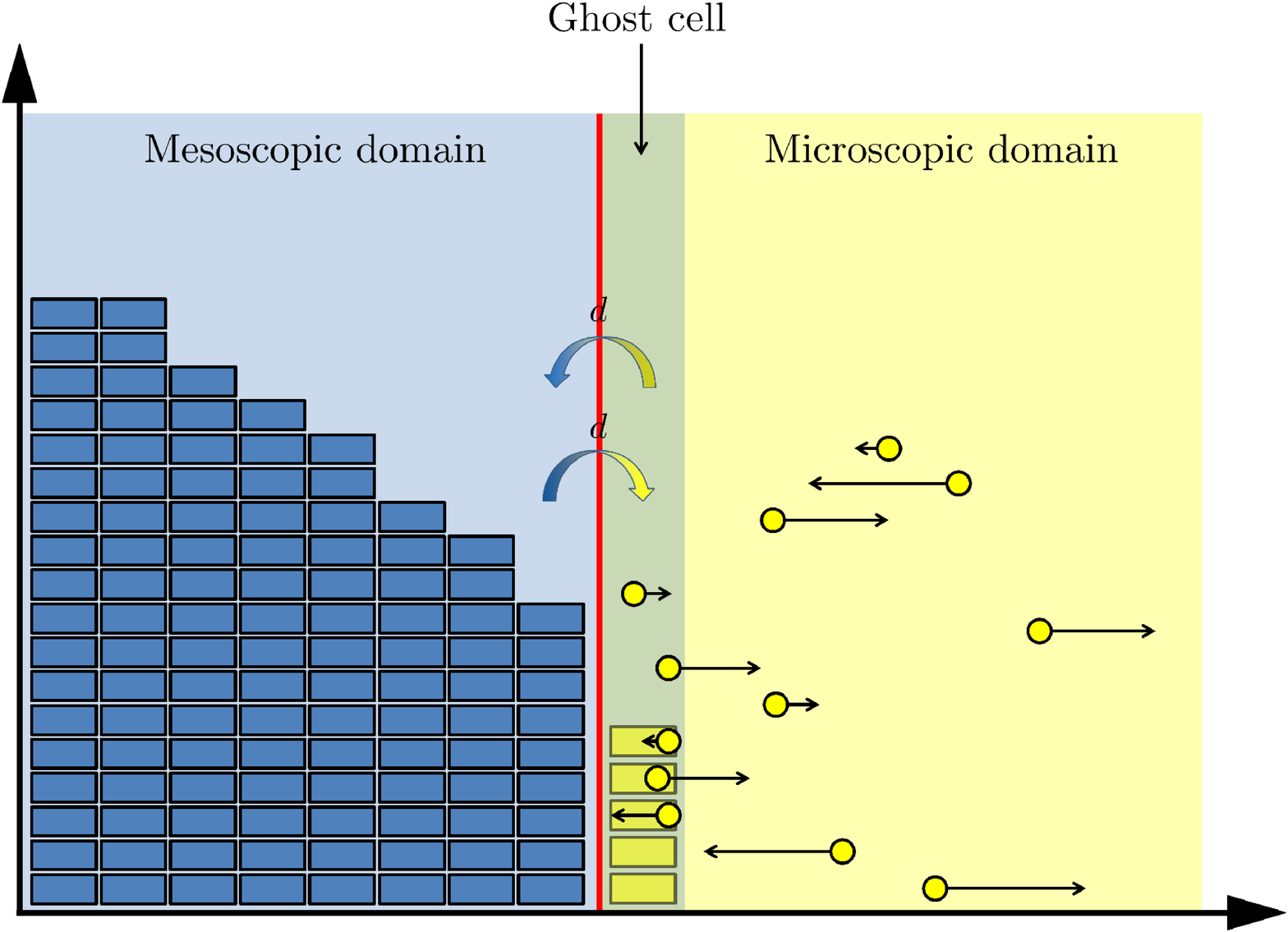}
\caption{Schematic for the GCM \citep{flegg2015cmc}. The blue boxes represent particles within each compartment and the yellow dots represent individual particles. These particles are shown with a volume, but in the simulations do not have a mass or volume. The particles reside on the one-dimensional line, but have been illustrated in the plane in order to show the directions and magnitudes of their next movement clearly (black arrows). The yellow boxes within the ghost-cell correspond to the number of Brownian particles which reside within it. The coloured arrows in the centre are similar to those in Figure \ref{fig:yates2015pcm}.}
\label{fig:flegg2015cmc}
\end{figure}

As an illustrative example for the mesoscopic-to-microscopic methods, we look at the ghost cell method (GCM), developed by \citet{flegg2015cmc}. The domain is divided into two subdomains, which we refer to as $\smallsub{\Omega}{C}$ and $\smallsub{\Omega}{B}$, within which the system is evolved according to a compartment-based method and Brownian dynamics respectively. As in the PCM (see Section \ref{subsection:macro-meso_hybrid_example}), $\smallsub{\Omega}{C}$ is split into $K$ compartments of width $h_c$, so that $|\smallsub{\Omega}{C}|=Kh_c$. In the Brownian subdomain, particles move in continuous space and a reflective boundary is enforced at the interface to prevent individual particles from entering the compartment-based region due to Brownian jumps. In order to allow the particles to move between the two subdomains, the authors construct a ``ghost cell'' in $\smallsub{\Omega}{B}$ adjacent to the interface with $\smallsub{\Omega}{C}$, which is the same width, $h_c$, as the compartments. We present a schematic for this method in Figure \ref{fig:flegg2015cmc}.

Particles move across the interface in both directions according to compartment-based dynamics, with the ghost cell constituting an extra compartment. In order to calculate the propensity function for particles to jump out of the ghost cell, the number of particles in that region of space is simply counted and multiplied by the compartment-based jump rate, $d$. The Brownian dynamics are implemented with a time-based algorithm and the compartment-based dynamics with an event-driven algorithm. At any time point, the time until the next compartment-based event (including jumps out of and into the ghost cell) is found according to \eqref{eqn:next_event_time}. It is then determined whether this event takes place before the next Brownian update. If a Brownian update comes first, the Brownian dynamics are evolved within $\smallsub{\Omega}{B}$ for a small time interval, $\Delta t$ according to \eqref{eqn:BD}. Otherwise, the mesoscopic event corresponding to the waiting time is determined and implemented. If a jump from the last compartment to the ghost-cell is enacted, a single particle is removed from the final compartment and is initialised with position chosen uniformly at random across the ghost cell. For movement across the interface in the opposite direction, one of the Brownian particles in the ghost-cell is chosen uniformly at random and removed from the system. An extra particle is then added to the final compartment of $\smallsub{\Omega}{C}$. Pseudocode for the GCM for diffusion only is provided in Algorithm \ref{alg:GCM}.

\begin{Algorithm}{Ghost cell method (diffusion only)} \label{alg:GCM}
\item Initialise time $t=t_0$, set the final time $T$. Specify the Brownian update step $\Deltat$ and set the next Brownian update time to be $t_\Delta = t_0 + \Deltat$.
\item Initialise particles in the compartments of $\smallsub{\Omega}{C}$ and Brownian particles in $\smallsub{\Omega}{B}$.
\item Calculate propensity functions for each compartment given by $\alpha_i(t)=dn_i(t)=Dn_i(t)/h_c^2$ for $i=1,\dots,K$, where $n_i(t)$ is the number of particles in compartment $i$ at time $t$. Calculate the propensity function for diffusion from the ghost cell, $\smallsub{\alpha}{GC}(t) = \smallsub{n}{GC}(t)D/h_c^2$, where $\smallsub{n}{GC}(t)$ is the number of particles in the ghost cell at time $t$. \label{item:GC_Loop}
\item Sum the propensity functions to find $\alpha_0(t)$.
\item Determine the time $\tau$ until the next compartment-based event according to equation \eqref{eqn:next_event_time}. Set $t_c = t+\tau$.
\item If $t_c\leq t_\Delta$, then the next compartment-based event occurs:
\begin{enumerate}[(a)]
\item Choose the event with probability proportional to the associated propensity function.
\item If the event corresponds to a diffusive jump out of the ghost-cell and into the last compartment, choose one particle in the ghost cell at random to remove and place it in the final compartment of $\smallsub{\Omega}{C}$.
\item If the event corresponds to a particle jumping from the final compartment of $\smallsub{\Omega}{C}$ to the ghost cell, remove a particle from the final compartment and add place it with position chosen uniformly at random across the width of the ghost cell.
\item If the event corresponds to a purely compartment-based event, implement the jump according to the usual compartment-based dynamics.
\item Update time $t = t_c$.
\end{enumerate}
\item If $t_\Delta < t_c$, we update the Brownian system:
\begin{enumerate}[(a)]
\item Update the positions of all particles using \eqref{eqn:next_event_time}.
\item Complete reactions using an appropriate method \citep{smoluchowski1917vem,andrews2004ssc,erban2009smr}.
\item Update time $t = t_\Delta$. Update $t_\Delta = t+t_\Delta$.
\end{enumerate}
\item If $t<T$, return to \ref{item:GC_Loop}, otherwise stop.
\end{Algorithm}

\begin{figure}[h!]
	\begin{center} 
	\subfigure[][]{
		\includegraphics[width=0.3\textwidth,trim={20pt 0pt 0pt 0pt},clip]{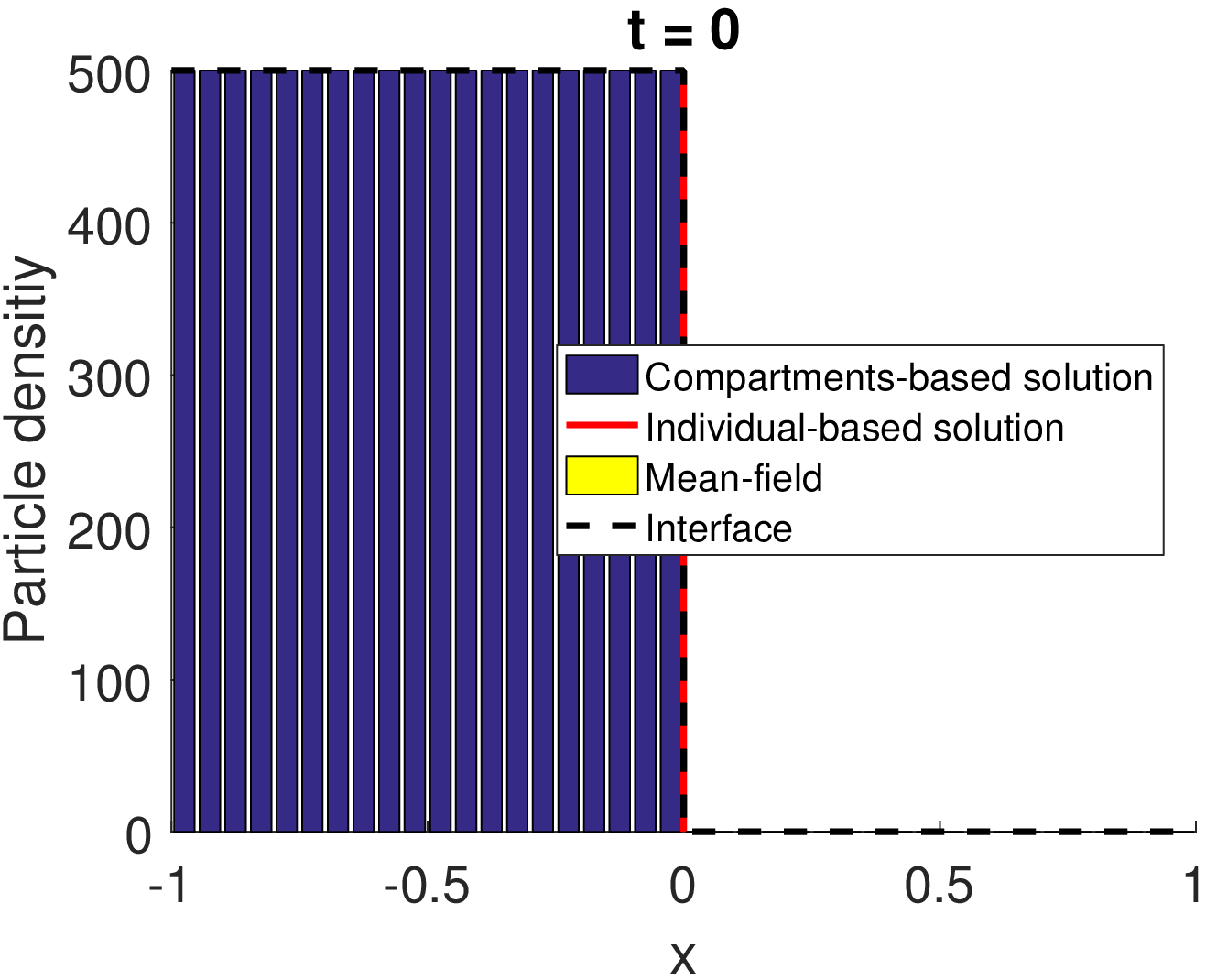}
		\label{fig:flegg2015cmc_1}
	}
	\subfigure[][]{
		\includegraphics[width=0.3\textwidth,trim={20pt 0pt 0pt 0pt},clip]{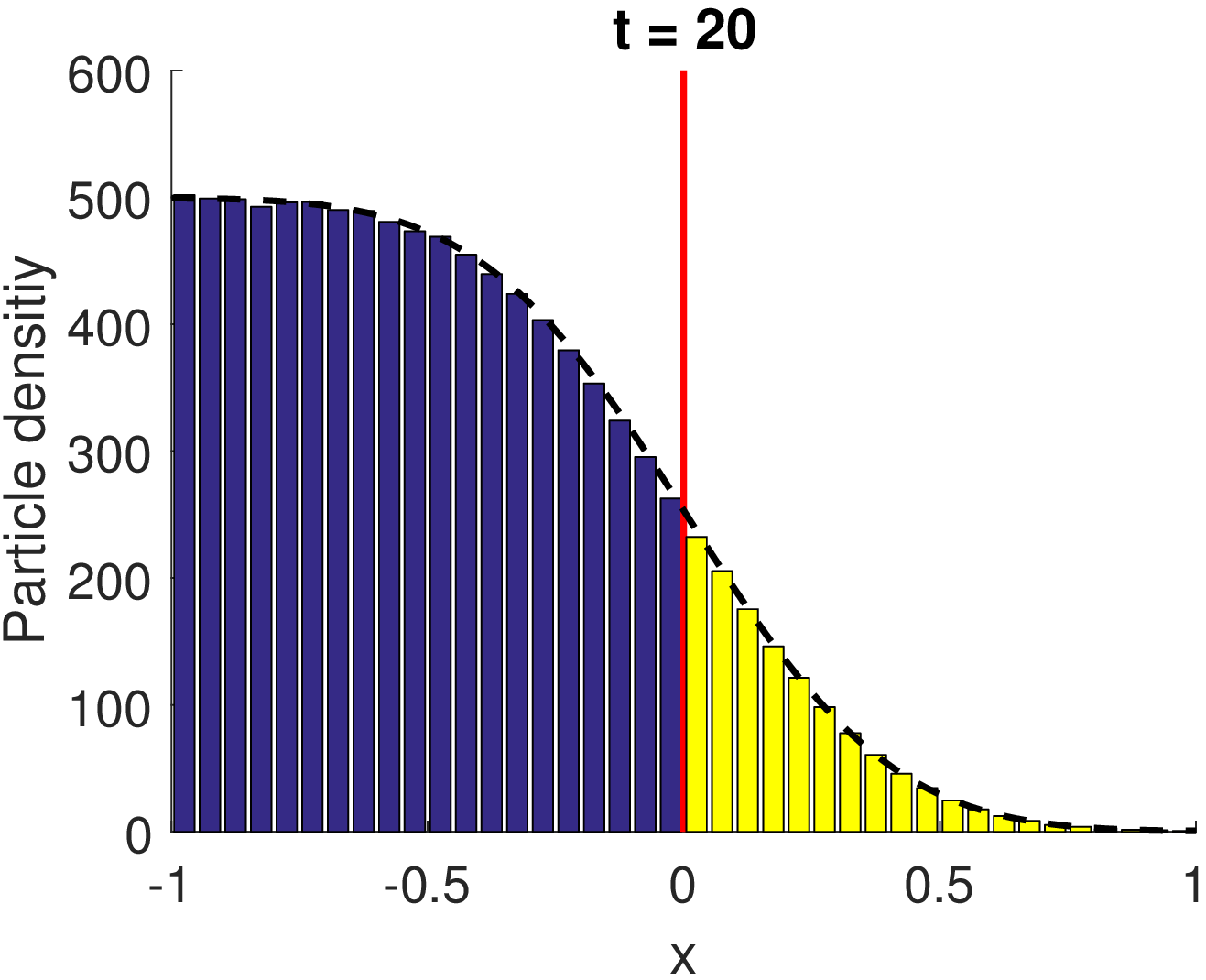}
		\label{fig:flegg2015cmc_2}
	}
	\subfigure[][]{
		\includegraphics[width=0.3\textwidth,trim={20pt 0pt 0pt 0pt},clip]{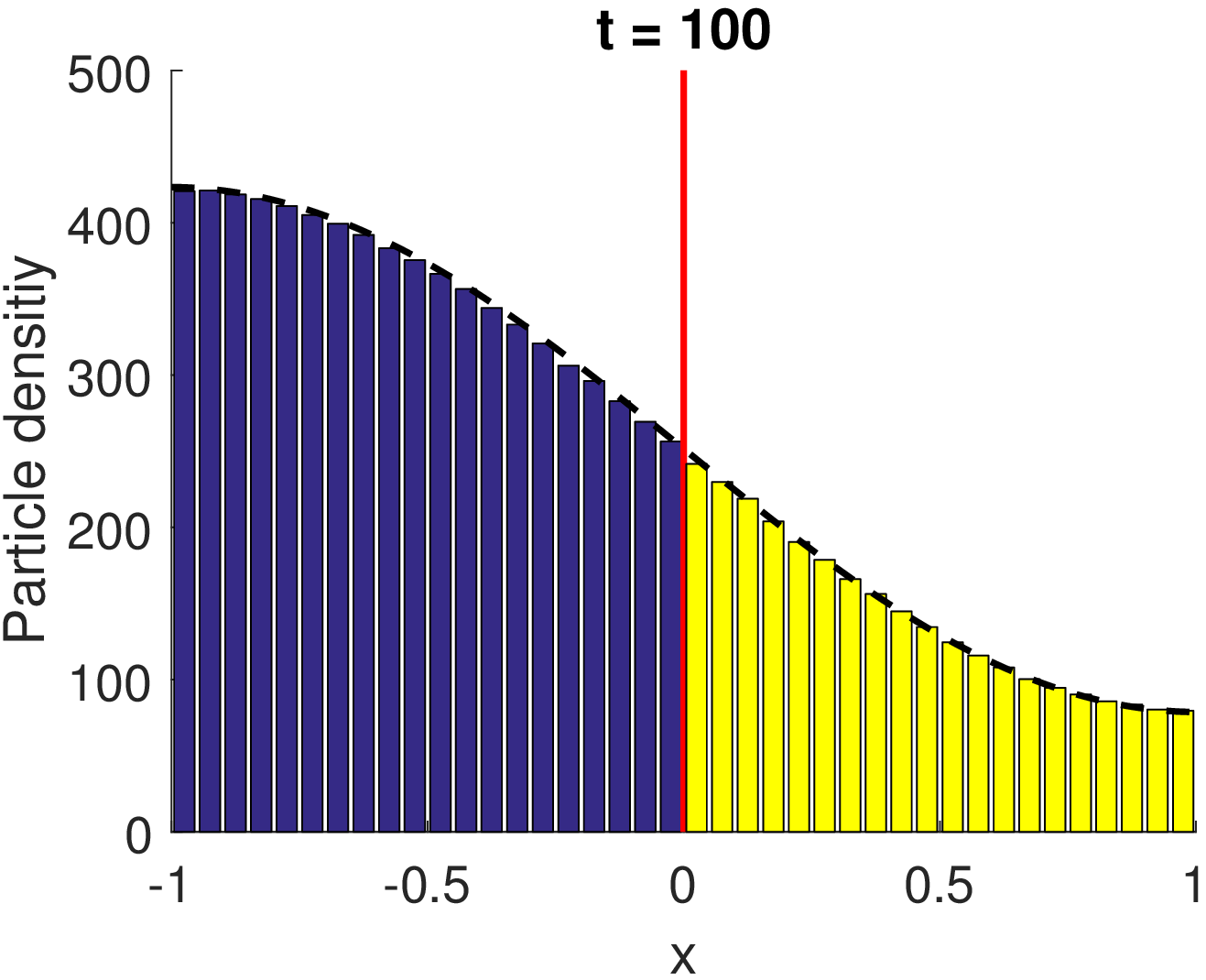}
		\label{fig:flegg2015cmc_3}
	}
	\caption{A replication of results from the GMC \citep{flegg2015cmc}. Descriptions are as in Figure \ref{fig:yates2015pcm_example}, with the addition that yellow bars denote the `binned' solution of the individual-based simulation in the hybrid method. Parameter values are as in the text.}
	\label{fig:flegg2015cmc_example}
\end{center}
\end{figure}

We have replicated some results from \citet{flegg2015cmc} using the GCM. These are displayed in Figure \ref{fig:flegg2015cmc_example}. As in the PCM, we have placed the interface centrally, $I=0$, with the mesoscopic subdomain at $\smallsub{\Omega}{C}=(-1,0)$ and the microscopic subdomain situated at $\smallsub{\Omega}{B}=(0,1)$. We set the Brownian update step to be $\Deltat=0.01$, and all other parameters are the same as the pseudo-compartment simulation.

\subsection{Other meso-micro hybrid methods} \label{section:other_meso-micro_hybrid_models}
% (I) flegg2012trm

We now outline the remaining meso-micro hybrid methods summarised in Table \ref{table:micro_meso_summary}. Many of these papers are variations of, or applications of, the same method, namely the two-regime method \citep{flegg2012trm}. We start by describing this method, and then follow by describing the adaptations and applications. We then consider two further methods, which fall under the operator splitting category \citep{klann2012hsg,hellander2012cmm}.

\begin{figure}[h!]
\centering
	\subfigure[][]{
		\includegraphics[width=0.48\textwidth, trim={15pt 15pt 15pt 15pt}, clip]{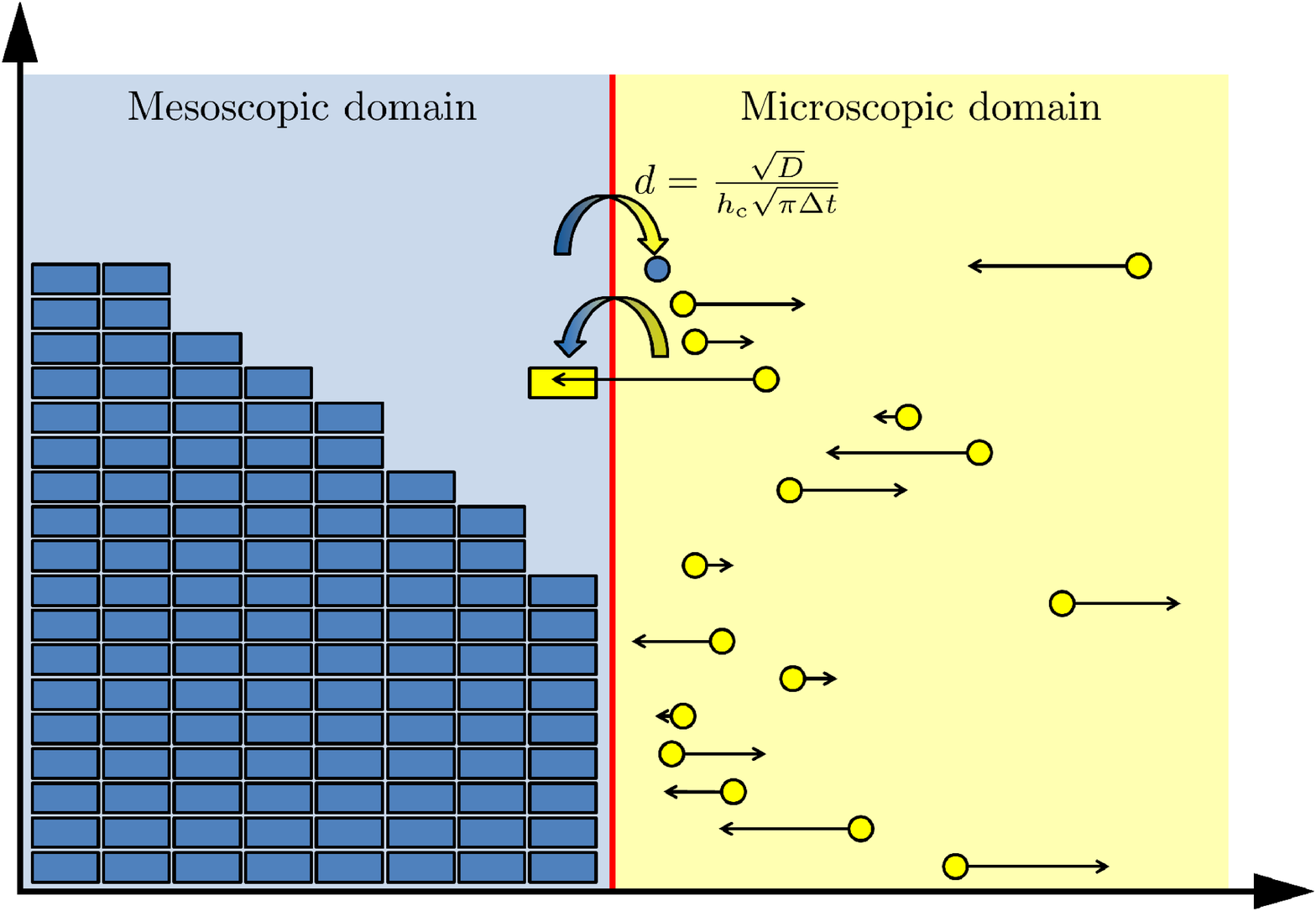}
		\label{fig:flegg2012trm}}
	\subfigure[][]{
		\includegraphics[width=0.48\textwidth, trim={15pt 15pt 15pt 15pt}, clip]{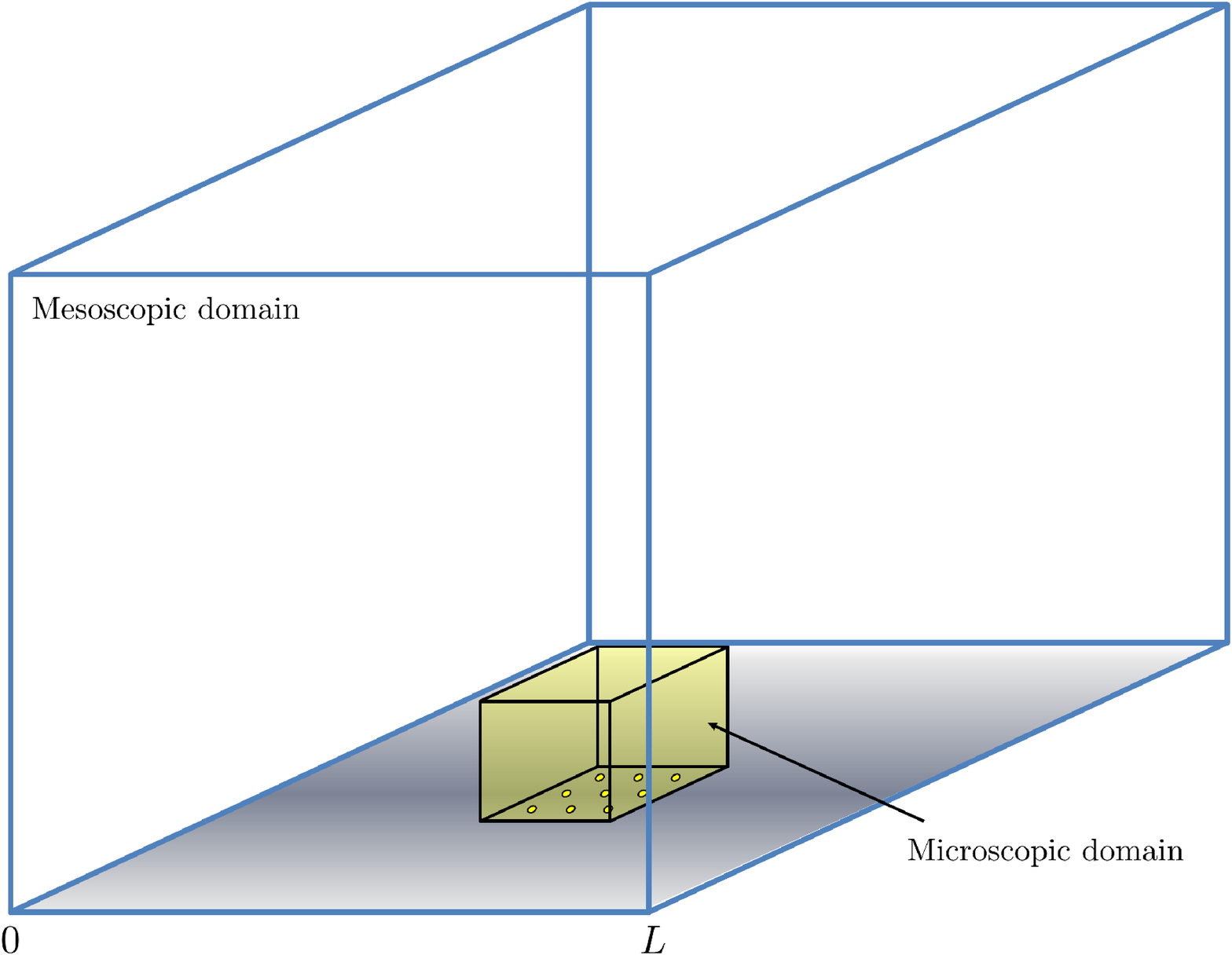}
		\label{fig:dobramysl2015pbd}}
\caption{\subref{fig:flegg2012trm} Schematic for the TRM \citep{flegg2012trm}. The blue blocks and yellow dots are as described in Figure \ref{fig:flegg2015cmc}. The arrow from left to right over the interface denotes the jump in this direction, with the specified altered jump rate. In this jump rate, $D$ is the macroscopic diffusion coefficient, $h_a$ is the width of a compartment and $\Delta t$ is the time-step used to evolve the particles in the Brownian-based subdomain. The other cross interface arrow represents jumps in the other direction. The yellow rectangle and blue particle near the interface represent particles converted from one modelling regime to the other upon crossing the interface in either direction according to the method described. \subref{fig:dobramysl2015pbd} Schematic for the application of the TRM to the problem of calcium-induced calcium release \citep{dobramysl2015pbd}. The blue outlined box denotes the outer boundaries of the compartment-based subdomain. All boundaries are absorbing, apart from the grey one (bottom), which is reflective. The yellow box in the centre of the lower face is the microscopic subdomain, containing nine ion channels (yellow circles). For simplicity, no particles or compartments are displayed in this schematic.}
\label{fig:TRMs}
\end{figure}

Some of the authors of the GCM previously developed the two-regime method (TRM) \citep{flegg2012trm} to couple compartment-based and Brownian-based dynamics. The individual particle paths are evolved according to independent Browninan motions, whilst the compartment regime is updated using the on-lattice, event-based next reaction method \citep{gibson2000ees}. Flux over the interface from the compartment-based subdomain to the Brownian-based subdomain is implemented using an altered jump rate to ensure that the flux over the interface is consistent with diffusion. If a particle is selected to jump across the interface from the final compartment to the Brownian-based subdomain, a particle is removed from the relevant compartment and placed at a position selected from a normalised error function probability distribution function. When a particle jumps from the microscopic subdomain to the mesoscopic subdomain, it is simply removed and added to the compartment it has moved in to. The TRM is represented schematically in Figure \ref{fig:TRMs} \subref{fig:flegg2012trm}.

% (I) robinson2014atr

\citet{robinson2014atr} introduce an extension to this method, called the adaptive TRM (ATRM), which adds an adaptive interface to the algorithm. The interface is moved in order to ensure that the subdomain that is to be simulated using the computationally intensive particle-based dynamics is as small as possible. The interface can only move in discrete steps, which are the same size as the width of a compartment in the mesoscopic subdomain. The interface movement condition is, similarly to \citet{moro2004hms} (see section \ref{section:other_macro-meso_hybrid_models}), a local condition. If the number of particles within a compartments width of the interface (and within the microscopic subdomain) is above a pre-specified level, the interface is moved into the microscopic subdomain, extending the mesoscopic subdomain. Conversely, if the number of particles in the compartment adjacent to the interface is below a distinct (lower) threshold, the interface moves towards the mesoscopic subdomain, increasing the size of the microscopic subdomain. The coupling between the compartment-based and Brownian-based methods is implemented exactly as the TRM \citep{flegg2012trm}.

% flegg2014atr

The TRM is generalised into two (and higher) dimensions by \citet{flegg2014atr}. The authors discuss in detail the case of a regular square lattice of points with a planar interface (in which the interface is either purely horizontal or vertical) and cases for which the interface may contain corners. The paper follows a similar method to the TRM paper, in which the authors calculate the factor by which the jump rate over the interface must be scaled by in order for a particle to move from the mesoscopic to microscopic subdomain, together with the rate in the opposite direction.

% (I) dobramysl2015pbd

These methods can be applied to biologically relevant scenarios such as the formation of calcium puffs in a range of eukaryotic cells \citep{dobramysl2015pbd,flegg2013dsn,erban2014msr}. \citet{dobramysl2015pbd} investigate the formations of such calcium puffs using the TRM. Calcium ions are modelled as diffusive particles, which can bind to activating and inhibiting receptors on the ion channels. Each channel contains four sub-channels, each with one activating and one inhibiting receptor. A sub-channel is activated if the activating receptor has a calcium ion bound to it, and the inhibiting one does not, and a channel is `open' if at least three of its four sub-channels are activated. When a channel is activated, a constant influx of particles is introduced into the domain. A particle can bind to a receptor with a given probability if it is within a small hemi-sphere of the receptor in question. Particles can also unbind. When particles unbind they are placed a given distance away from the receptor with a second probability. The authors simulate this process in a (three-dimensional) cube representing some part of the cytoplasm of the cell. One face of the cube represents part of the surface of the impermeable endoplasmic reticulum (the cell's major calcium store) upon which a reflecting boundary condition is implemented. In the centre of one of this faces are nine ion channels. On all other faces, an absorbing boundary condition is used. The authors couple the microscopic Brownian dynamics for particle motion in a small cube around the nine ion channels to a mesoscopic compartment-based regime throughout the rest of the domain. The mesoscopic regime is simulated using the next reaction method \citep{gibson2000ees}. This hybrid representation is used to investigate calcium puffs which occur when a calcium channel opens and then closes quickly, allowing for a large number of ions to enter the domain over a short time period. This problem is a good example of the need for hybrid methods to couple simulation methods at different scales. If this is simulated using a fully individual-based model, the computational complexity would be too high to simulate accurately within a reasonable time-frame.

% hellander2012cmm

Another method which falls into the meso-micro category is presented by \citet{hellander2012cmm}. This is an operator splitting method rather than a spatially-coupled hybrid method. The spatial domain is divided into discrete voxels and the algorithm allows for particular voxels or species to be described as either mesoscopic or microscopic.
%  The authors apply their algorithm to curved surfaces by taking local planar approximations and simulating movement on these proxy surfaces. 
The algorithm progresses using a splitting scheme. First the microscopic particles are frozen and the mesoscopic particles are progressed using the SSA \citep{gillespie1977ess}. Then, the mesoscopic particles are frozen to allow the microscopic particles to advance according to the Green's function reaction dynamics \citep{van2005gfr}. Finally, reactions between mesoscopic and microscopic particles are completed according to the microscopic algorithm, with an adjusted reaction rate to account for the difference in representation.

% klann2012hsg

Operator splitting is also employed by \citet{klann2012hsg}. The spatial domain (assumed three-dimensional) is split into equally sized cubic compartments. Within each of these subvolumes, some species are chosen to be simulated via the compartment-based paradigm using Gillespie's SSA, whilst others are evolved using the Brownian-based approach with a fixed time-step. Thus, different modelling paradigms are used for different species within the same voxel, but also potentially for the same species in different regions of the domain. For each species simulated under the compartment-based paradigm, a minimum time until the next occurrence of \textit{any} type of first-order reaction affecting that species (other than diffusive jumps) is stored. If a particle diffusively jumps out of a compartment (either into a region in which the compartment-based paradigm is being employed for that species or a region in which that species is being modelled as particles) then with probability inversely proportional to the number of particles of its species in the compartment it has just left, the jumping particle takes this minimum first order reaction time with it to the new compartment. The authors use an updated next reaction method (introduced by \citet{anderson2007mnr}) to implement both reactions and diffusive jumps for particles which are modelled using the compartment-based approach. For particles which are modelled microscopically, diffusion is completed via a discretised SDE which represents Brownian motion, while bimolecular reactions are simulated using the $\lambda$-$\rho$ methodology \citep{erban2009smr,lipkova2011abd}.

If an entire compartment changes description from mesoscopic to microscopic according to the specified criteria, the appropriate number of particles are initialised uniformly throughout the compartment. Of the new individual particles, one inherits the next reaction time for first order reactions from the mesoscopic description, whilst  exponentially distributed first reaction times which are later than the inherited time are generated for the others. For a conversion in the opposite direction, the next firing times for diffusive and second- (and higher-, if required) order reactions are calculated according to the standard Gillespie method. For first order reactions, the minimum time (over all the particles of the same species) is used. A similar mechanism is employed if only certain species change their description based on a threshold.

The number of unique methods that we have considered in this category is relatively small. However, the development of the TRM that we have reviewed, serves to demonstrate how a basic method can be altered to incorporate adaptive interfaces and higher dimensions, as well as applied to genuinely multiscale problems. In the following section, we investigate a third category of our spatial coupling involving macroscopic and microscopic models.

% To read:
% fange2010srd

\section{Macroscopic-to-microscopic methods} \label{sect:MacroMicro}

In this section, we will introduce and review models which couple macroscopic dynamics to microscopic dynamics, which we will refer to as ``macro-micro'' hybrid methods. We list and describe the macro-micro hybrid methods covered in this section in Table \ref{table:macro_micro_summary}. We begin by summarising an illustrative example of a macro-micro hybrid method, the auxiliary region method (ARM) \citep{smith2017arm} and present pseudocode for its implementation. We then summarise other existing macro-micro hybrid methods.

Hybrid methods that couple the macroscopic continuum representations to discrete microscopic dynamics have been relatively poorly studied in comparison to macro-meso and meso-micro hybrid methods. One contributing factor is the fact that such hybrid algorithms bypass the intermediate mesoscale representations of particle dynamics, meaning that the scale separation gap which they must bridge is greater than either of the other two hybrid paradigms. Primarily though, we postulate the relative dearth of  macro-micro hybrid methods is due to the inherent difficulty when converting individual Brownian particles into continuum mass (and vice-versa) when coupling individual-based microscopic methods to continuum macroscopic continuum representations.

Although they are less common, macroscopic-to-microscopic methods provide useful insight into a number of biological and physical phenomena, such as the movement of cytochrome c particles in the presence of a charged surface \citep{gorba2004bds}.

\begin{center}
\begin{table}
\centering
\begin{tabular}{|m{0.2\textwidth} || m{0.3\textwidth} | m{0.2\textwidth} |}
\hline
\textbf{Paper} & \textbf{Type} & \textbf{System modelled}\\
\hline
\citet{smith2017arm} & Spatially-coupled,\linebreak non-adaptive,\linebreak no overlap & Reaction-diffusion \\
\hline
\citet{franz2012mrd} & Spatially-coupled,\linebreak non-adaptive,\linebreak no overlap/overlap & Reaction-diffusion \\
\hline
\citet{geyer2004ibd} & Spatially-coupled,\linebreak non-adaptive,\linebreak no overlap & Reaction-diffusion \\
\hline
\citet{gorba2004bds} & Spatially-coupled,\linebreak non-adaptive,\linebreak no overlap & Electrostatics\\
\hline
\citet{alexander2002ars} & Spatially-coupled,\linebreak non-adaptive,\linebreak no overlap & Reaction-diffusion \\
\hline
\citet{alexander2005ars} & Spatially-coupled,\linebreak non-adaptive,\linebreak no overlap & Viscous gas (train model) \\
\hline
\citet{plapp2000mrw} & Spatially-coupled,\linebreak non-adaptive,\linebreak no overlap & Dendritic growth \\
\hline
\end{tabular}
\caption{A summary of the macro-micro hybrid papers that will be covered in this section. The methods in the macro-micro hybrid papers are designed for modelling a diverse array of applications. Each of these papers are concerned with the development of a novel hybrd method, apart from the paper by \citet{gorba2004bds}, which uses method they previously developed \citep{geyer2004ibd} in order to model the movement of cytochrome c molecules in the presence of a charged surface. Descriptors are as in Table \ref{table:macro_meso_summary}.}
\label{table:macro_micro_summary}
\end{table}
\end{center}

% (I) ARM <<<<<<

\subsection{Illustrative example of a macro-micro hybrid -- the auxiliary region method}

\begin{figure}[h!]
\centering
\includegraphics[width=0.8\textwidth]{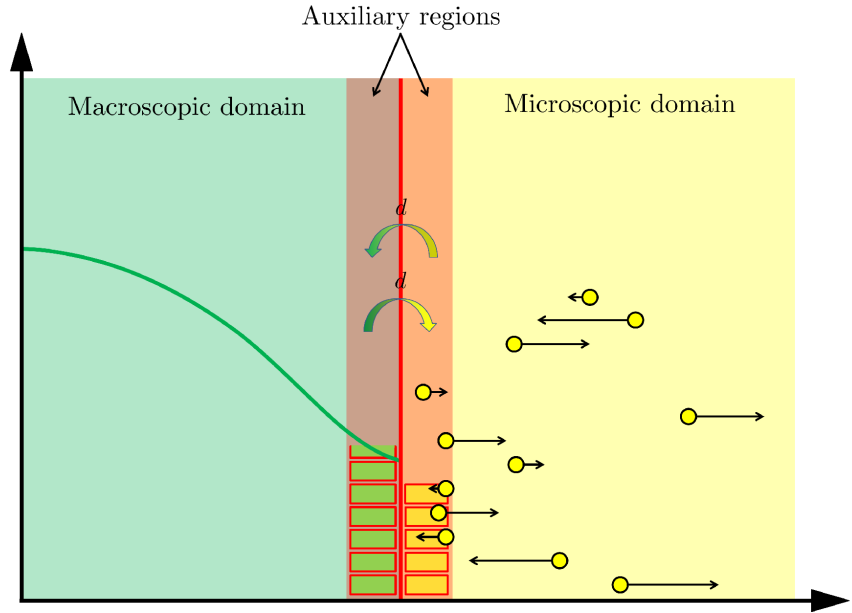}
\caption{Schematic for the ARM \citep{smith2017arm}. The green line and yellow dots represent the same phenomena as in Figures \ref{fig:yates2015pcm} and \ref{fig:flegg2015cmc} respectively. The auxiliary regions on either side of the interface are highlighted in red. The green and yellow boxes within auxiliary regions represent compartment-based particle numbers in the PDE and Brownian auxiliary regions respectively. The coloured arrows in the centre represent the conversion of particles between the mesoscopic and microscopic auxiliary regions, similar to those in Figure \ref{fig:yates2015pcm}.}
\label{fig:ARM_Schematic}
\end{figure}

As an illustrative example of a macroscopic-to-microscopic hybrid method, we consider the auxiliary region method (ARM) \citep{smith2017arm}. The ARM couples a PDE for reaction-diffusion systems in a subdomain $\smallsub{\Omega}{P}$ to individual-based Brownian dynamics in a subdomain $\smallsub{\Omega}{B}$. Both of the subdomains have zero flux boundaries at the interface so that no PDE mass ``leaks'' into the individual-based subdomain, and vice versa. Flux over the interface is governed strictly by compartment-based dynamics between the two auxiliary regions, $\smallsub{\Omega}{PA}$ and $\smallsub{\Omega}{BA}$, adjacent to the interface within the PDE and Brownian subdomains respectively. The one-dimensional schematic for the ARM is displayed in Figure \ref{fig:ARM_Schematic}.

In order to implement compartment-based jumps over the interface, particle numbers within each of the auxiliary regions are calculated. For the PDE auxiliary region, the number of auxiliary particles can be calculated as:
\begin{equation}
\smallsub{n}{PA}(t) = \displaystyle\int_{\smallsub{\Omega}{PA}}{c(x,t)\ dx},
\label{eqn:particle_numbers_PDE_AR}
\end{equation} 
where $c(x,t)$ is the solution to the hybrid PDE in $\smallsub{\Omega}{P}$. Similarly, the number of particles within the Brownian auxiliary region is \begin{equation}
\smallsub{n}{BA}(t) = \left|\set{j}{y_j(t)\in \smallsub{\Omega}{BA}}\right|,
\label{eqn:particle_numbers_BD_AR}
\end{equation} 
with $y_j(t)$ the position of particle $j$ at time $t$. These auxiliary particle numbers are used to calculate propensity functions, which are then employed in an event-driven SSA which determines the time of the next jump across the interface. These auxiliary regions, the dynamics of which are simulated using the compartment-based method, are designed to bridge the gap between the finest and coarsest representations. Particles which jump from the macroscopic subdomain to the microscopic subdomain are removed from the PDE auxiliary region $\smallsub{\Omega}{PA}$ by removing one particle's worth of mass uniformly over its width, and are then initialised with position chosen uniformly at random within $\smallsub{\Omega}{BA}$, the Brownian auxiliary region. A movement in the opposite direction is completed by first choosing a particle in $\smallsub{\Omega}{BA}$ uniformly at random, removing it, and then adding a particle's worth of mass  to the PDE solution uniformly over the region $\smallsub{\Omega}{PA}$.

Reactions are completed using the appropriate methodology for the subdomain in which they reside, with the exception that for reactions with at least one set of participating particles lying within the Brownian auxiliary region, $\smallsub{\Omega}{BA}$. Firings of the reactions involving these subsets of particles are implemented according to the SSA in order to prevent the potential creation of individual-based particles within the PDE subdomain. Pseudocode for the implementation of the ARM is given in Algorithm \ref{alg:ARM}. For simplicity, we present the algorithm for a single species in one dimension.

\begin{Algorithm}{Auxiliary region method (ARM)}\label{alg:ARM} 
\item Initialise time $t=t_0$, set final time $T$, PDE/Brownian update time-step, $\Deltat$, the PDE discretisation grid size, $h_p$, and the auxiliary region width, $h_a$. Initialise particles in the PDE subdomain, $\smallsub{\Omega}{P}$, and the Brownian subdomain, $\smallsub{\Omega}{B}$, as required. Calculate the time until the next PDE and Brownian update step $t_\Delta = t+\Deltat$.

\item Calculate the number of particles $\smallsub{n}{PA}$ and $\smallsub{n}{BA}$ in the auxiliary regions, using formulae \eqref{eqn:particle_numbers_PDE_AR} and \eqref{eqn:particle_numbers_BD_AR} respectively. Consequently, calculate the corresponding propensity functions, $\smallsub{\alpha}{P}(t) = d\smallsub{n}{PA}(t)$ and  $\smallsub{\alpha}{B}(t)=d\smallsub{n}{BA}(t)$. Calculate propensity functions for any relevant reactions within $\smallsub{\Omega}{BA}$, and finally the sum of all the propensity functions to give $\alpha_0$. \label{item:while_loop_arm}

\item Calculate the time, $\tau$, until the next auxiliary region event according to equation (\ref{eqn:next_event_time}). Update the auxiliary region time $t_c = t+\tau$.

\item If $t_c<t_\Delta$
	\begin{enumerate}[(i)]
	\item Draw three random numbers $u_1,u_2,u_3\sim \text{Unif}(0,1)$.
	\item If $u_1\alpha_0(t) < \smallsub{\alpha}{PA}(t)$ (corresponding to a jump from $\smallsub{\Omega}{PA}$ to $\smallsub{\Omega}{BA}$):
		\begin{itemize}
		\item Remove a particle from the PDE auxiliary region according to $$c(x,t) = c(x,t) - \frac{1}{h_a}\mathds{1}_{[x\in\smallsub{\Omega}{PA}]}.$$
		\item Initialise a new particle of uniformly within $\smallsub{\Omega}{BA}$ with position $y^*=u_3h_a+I$.
		\end{itemize}
		Else if $u_1\alpha_0(t) < \smallsub{\alpha}{P}(t)+\smallsub{\alpha}{B}(t)$ (corresponding to a jump from $\smallsub{\Omega}{BA}$ to $\smallsub{\Omega}{PA}$):
		\begin{itemize}
		\item Choose a particle at random from within the Brownian auxiliary region and remove it from the system by selecting an index $q$ according to $q=\lceil u_3\smallsub{n}{BA}\rceil$ (where $\lceil x\rceil$ represents the smallest integer greater than $x$).
		\item Add a new particle into the PDE auxiliary region according to $$c(x,t) = c(x,t) + \frac{1}{h_a}\mathds{1}_{[x\in\smallsub{\Omega}{PA}]}.$$
		\end{itemize}
		Else (corresponding to a reaction in $\smallsub{\Omega}{BA}$)
		\begin{itemize}
		\item Use $u_2$ to choose a reaction to be implemented from the list of possible reactions with probability proportional to its propensity function.
		\item Enact the reaction chosen in the previous step according to the usual kinetics of the reaction pathway 
		\end{itemize}
	\item Set $t = t_c$
% 	and update particle numbers $\smallsub{n}{P}^k$ and $\smallsub{N}{HB}^k$ for the species $k$.
	\end{enumerate}
	Else 
	\begin{enumerate}[(i)]
	\item Update the PDE system using an appropriate numerical method.
	\item Implement any reactions in $\smallsub{\Omega}{B}$ using any appropriate method. Note that production reactions should be implemented after any degradation reactions in order to prevent particles being created and destroyed in the same time-step.
	\item Update the positions of the Brownian particles according to equation \eqref{eqn:BD}, including any boundary conditions
	\item Set $t = t_\Delta$, update $t_{\Delta} = t + \Deltat$.
	\end{enumerate}
\item If $t<T$, return to \ref{item:while_loop_arm}, otherwise stop.
\end{Algorithm}

As with the PCM and GCM, we have replicated some of the results from \citep{smith2017arm} using the ARM. For these examples, the macroscopic subdomain is $\smallsub{\Omega}{P}=(-1,0)$ and the microscopic, Brownian subdomain is $\smallsub{\Omega}{B}=(0,1)$. Both auxiliary regions are set to be size $h_a= 0.05$, and the time-step for both the Brownian and PDE updates are set to $\Deltat = 0.01$. All other parameter values are as in the previous simulations. The results are shown for the same initial condition as in Figure \ref{fig:flegg2015cmc_example}.

\begin{figure}[h!]
	\begin{center} 
	\subfigure[][]{
		\includegraphics[width=0.3\textwidth,trim={20pt 0pt 0pt 0pt},clip]{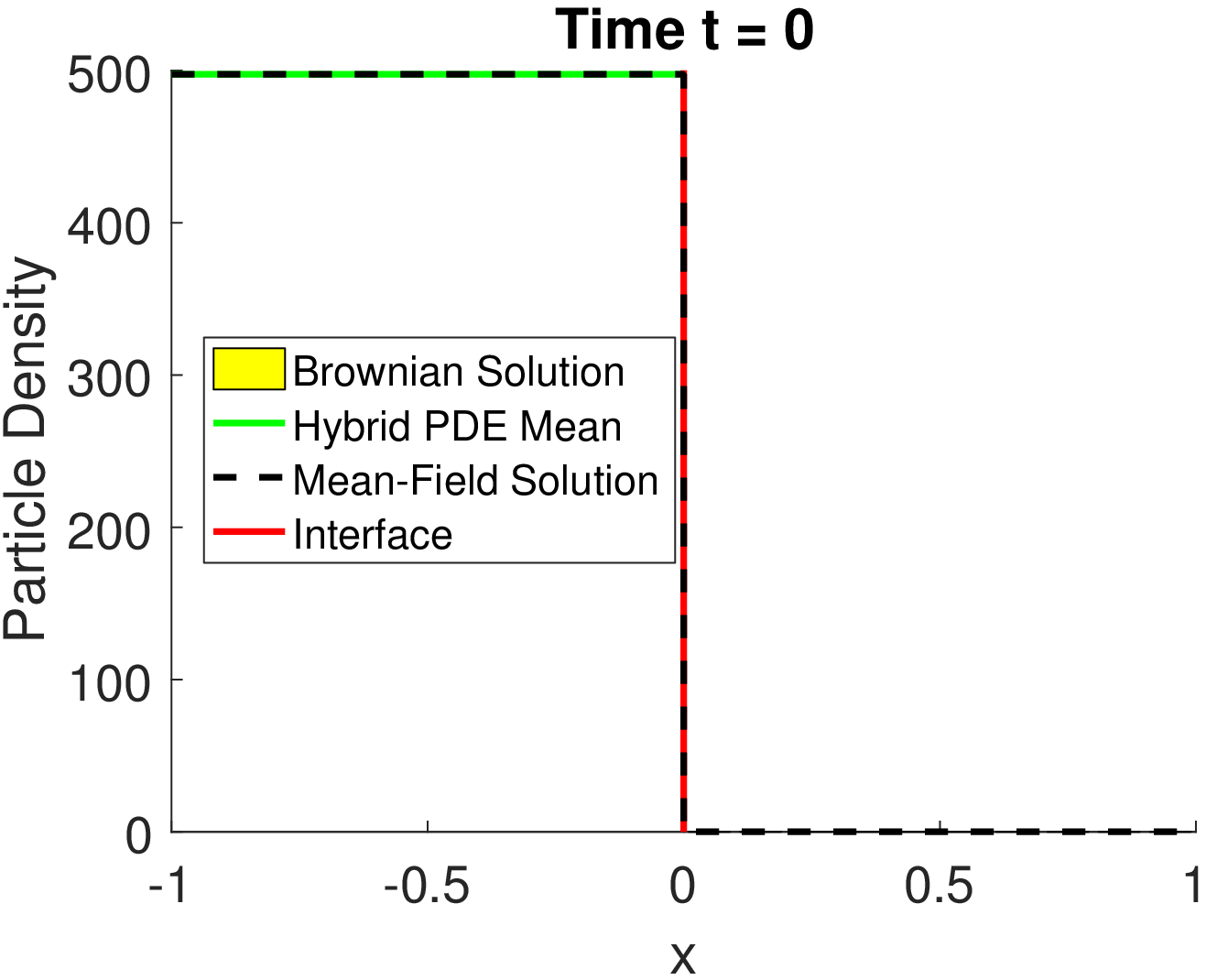}
		\label{fig:ARM_1}
	}
	\subfigure[][]{
		\includegraphics[width=0.3\textwidth,trim={20pt 0pt 0pt 0pt},clip]{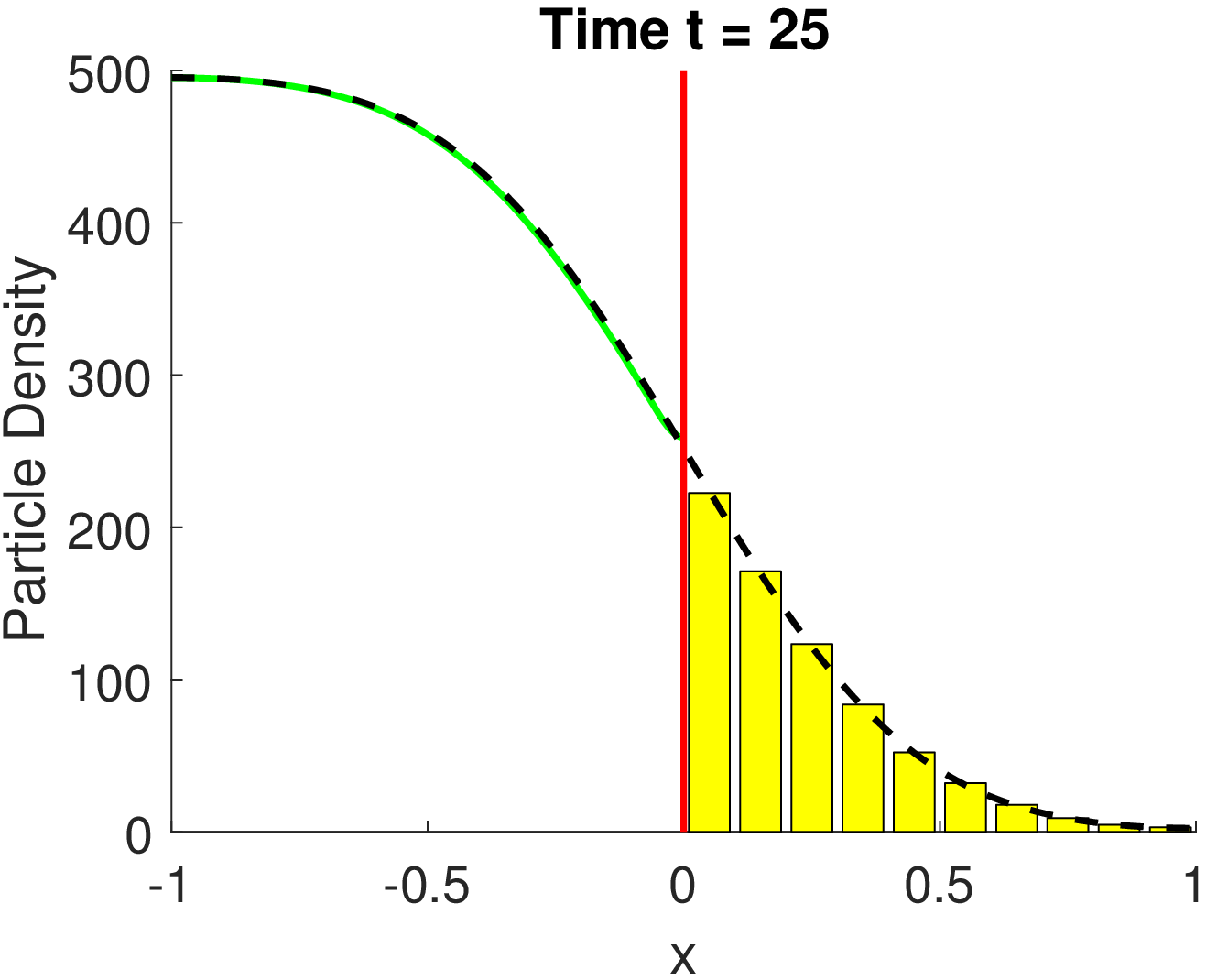}
		\label{fig:ARM_2}
	}
	\subfigure[][]{
		\includegraphics[width=0.3\textwidth,trim={20pt 0pt 0pt 0pt},clip]{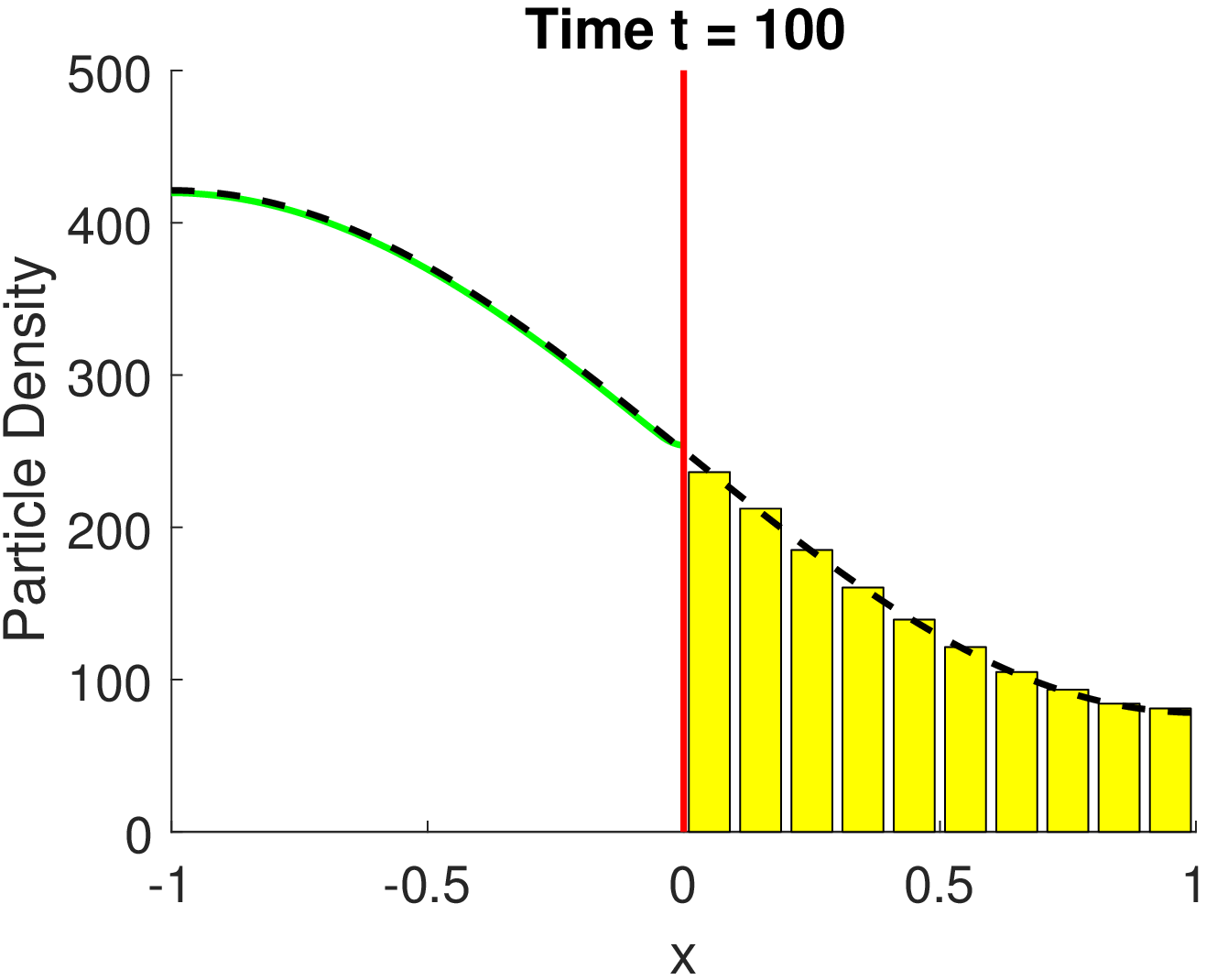}
		\label{fig:ARM_3}
	}
	\caption{Replication of results from the ARM \citep{smith2017arm}. Descriptions for the PDE and Brownian domains are as in Figures \ref{fig:yates2015pcm_example} and \ref{fig:flegg2015cmc_example}, respectively, with parameter values given in the text.}
	\label{fig:ARM_example}
\end{center}
\end{figure}

% (I) franz2013mrd

\subsection{Other macro-micro hybrid methods}

\begin{figure}[h!]
\centering
\includegraphics[width=0.8\textwidth]{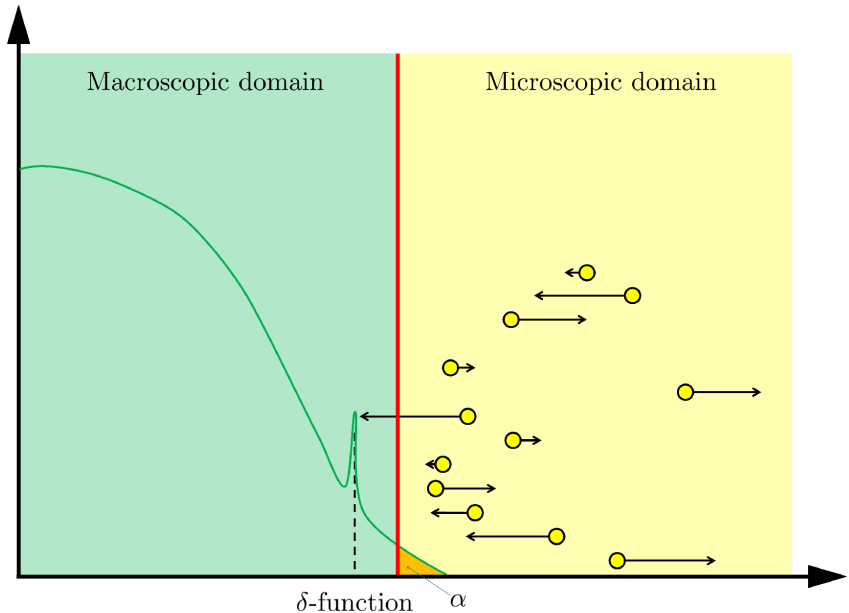}
\caption{Schematic for the method by \citet{franz2012mrd} (without overlap region). The green line and yellow dots represent the same quantities as in Figure \ref{fig:ARM_Schematic}. The orange mass labelled $\alpha$ is the amount of mass that flows over the interface in a small time-interval (comprising several PDE updates). Its total mass is used to find the probability of a particle being initialised in the microscopic subdomain, and its profile acts as a scaled probability density function for the position of the new molecule. The spike in the PDE solution is representative of a Dirac delta-function which is added to the PDE at the location that a Brownian particle has jumped to from the Brownian subdomain.}
\label{fig:franz2013mrd}
\end{figure}

\citet{franz2012mrd} present a macro-micro hybrid method in which the coupling is completed directly, without the use of a compartment-based intermediary regime (Figure \ref{fig:franz2013mrd}). In the microscopic subdomain, particles evolve their positions according to Brownian motion. The corresponding Fokker-Planck equation which describes the evolution of the probability density of each particle is the diffusion equation. 
% By coupling the dynamics of the individual particle to a scaled version of the diffusion equation via an interface which separates the two subdomains, \citet{franz2012mrd} reduce he necessity for the computationally expensive simulation of individual particles and replace it with the fixed cost of numerical solution of a PDE. 

The conversion of PDE mass to individual particles is achieved by allowing PDE mass to flow over the interface and probabilistically determining whether sufficient mass has crossed the interface to warrant the instantiation of a new Brownian particle. Conversely, Brownian particles crossing he interface in the opposite direction are realised as delta function contributions to the PDE solution at the position at which they arrive at the end of their jump (\ref{fig:franz2013mrd}).

Upon finding that their initial coupling algorithm can correctly maintain mean particle concentrations, but incorrectly matches particle variance profiles, \citet{franz2012mrd} adapt their algorithm by incorporating an overlap region in which some of the mass is represented as PDE and some as Brownian particles. At the interface at one end of the overlap region, PDE mass is converted in to particles, as before, and at the the other end, particles are incorporated into the PDE by the addition of delta functions as previously. The addition of this overlap region corrects the variance of the particles in the purely Brownian region of the hybrid simulations. 

% (I) geyer2004ibd
% (I) gorba2004bds

\citet{geyer2004ibd} also allow mass from the PDE to flow over the interface. They introduce two methods to interface Brownian dynamics simulations for diffusion to a deterministic macroscopic density-based representation. The first method couples individual particles to a constant density reservoir, whereas in the second, the macroscopic subdomain itself evolves according to a discretised version of the diffusion equation. In the first case, the authors ensure the correct movement over the boundary by removing particles when they cross into the reservoir from the Brownian dynamics subdomain, and inserting new particles into the Brownian dynamics subdomain with an appropriate rate and position. The rate and position are determined by using the fundamental solution of the diffusion equation to calculate the probability density function (PDF) and magnitude of mass which has flowed over the interface in the intervening time period. This can then be used to determine if, and where, a particle should be placed in the microscopic Brownian dynamics subdomain.

\begin{figure}[h!]
\centering
	\subfigure[][]{
		\includegraphics[width=0.45\textwidth]{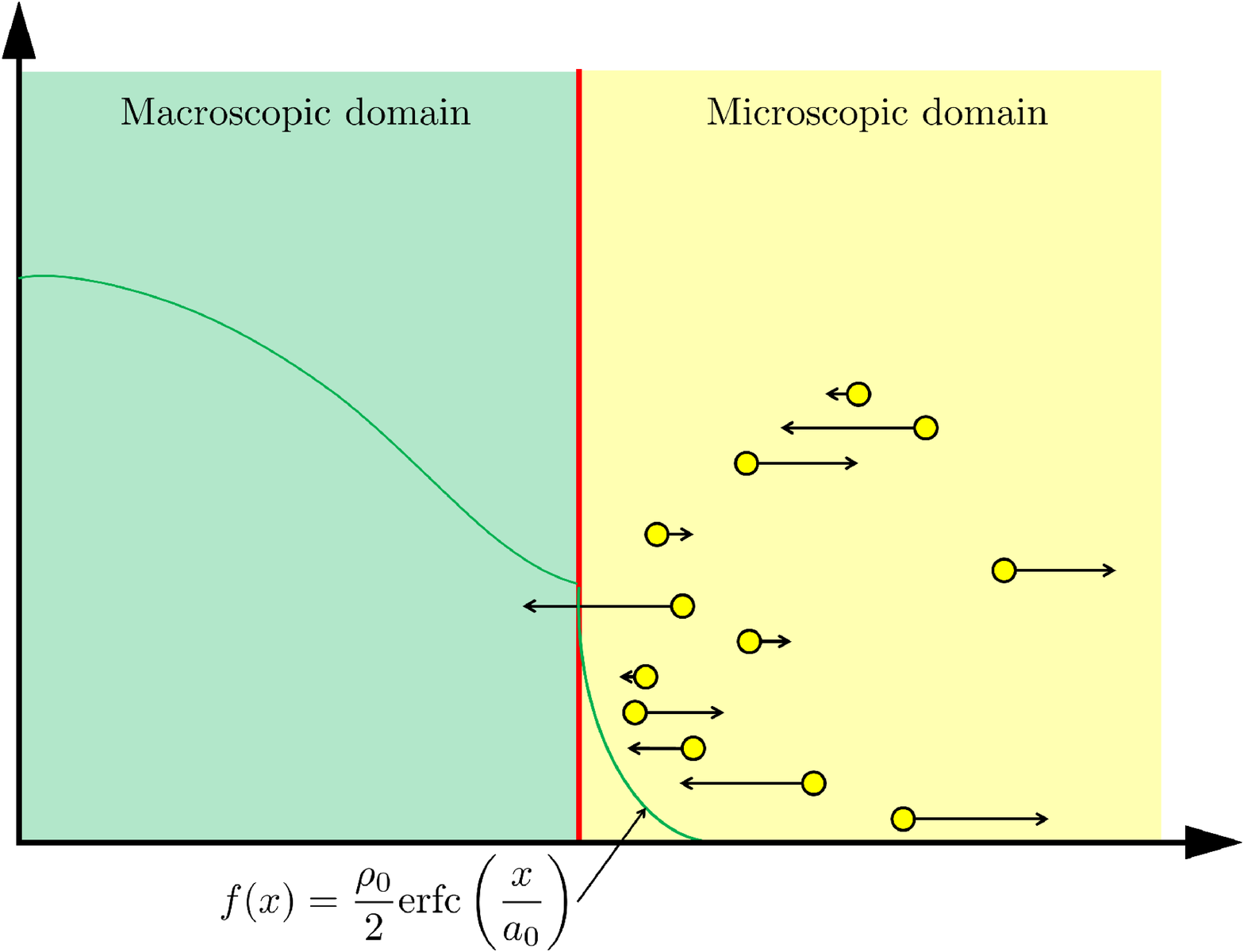}
		\label{fig:geyer2004ibd}}
	\subfigure[][]{
		\includegraphics[width=0.45\textwidth]{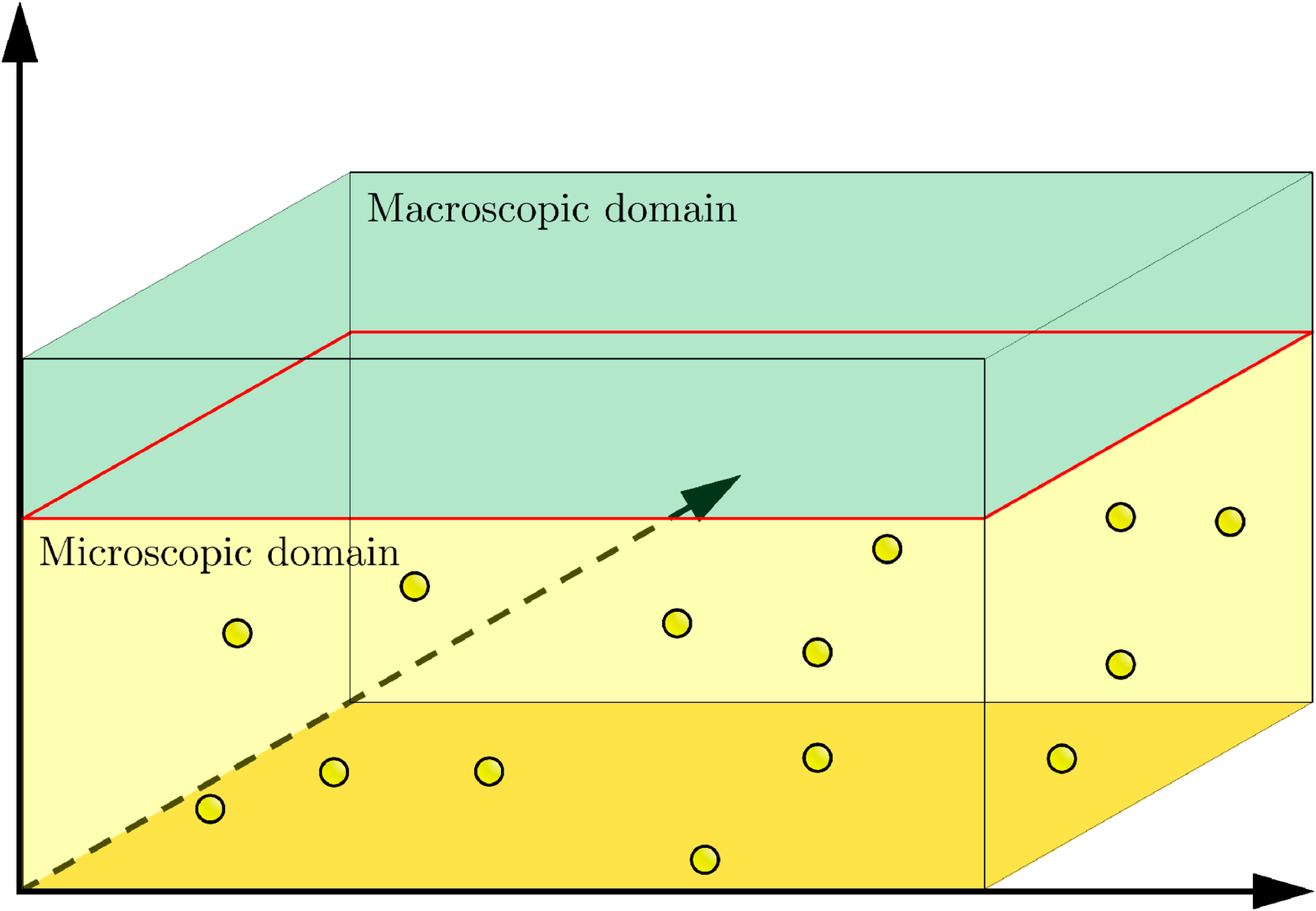}
		\label{fig:gorba2004bds}}
\caption{\subref{fig:geyer2004ibd} Schematic for the method presented by \citet{geyer2004ibd}. The green lines and yellow dots represent the same phenomena as in Figure \ref{fig:ARM_Schematic}. The additional green line which resides in the microscopic subdomain is the mass which flows over the interface after a given time, where $\rho_0$ is the density at the PDE meshpoint adjacent to the interface and $\sigma=2\sqrt{D\Deltat}$ is the average Brownian step size during a time interval of length $\Deltat$. \subref{fig:gorba2004bds} Schematic for the application presented by \citet{gorba2004bds}. The yellow dots are the same as in Figure \ref{fig:ARM_Schematic}, while the blue region is a constant density heat-bath. There are reflective boundary conditions on all sides of the computational domain, with the exception of the lower boundary, denoted in orange. This is a repulsive boundary caused by the van der Waals forces, representing the charged boundary.}  
\label{figure:geyer_and_gorba}
\end{figure}

For their second hybrid method (see Figure \ref{figure:geyer_and_gorba} \subref{fig:geyer2004ibd}), which couples Brownian particles to a dynamic PDE, the PDE mesh-point located closest to the interface is used to determine the probability density function of particles flowing into the Brownian subdomain (i.e. it is treated as a constant density reservoir as in the fixed density case). This relies on choosing the PDE mesh width sufficiently large (and thus sacrificing accuracy for the PDE solution) or the time step to be sufficiently small so that the majority of the mass that flows in to the Brownian subdomain originates in this region. However, the value of the PDE solution at this mesh-point is allowed to evolve dynamically according to diffusive fluxes. The flux into this PDE mesh-point from the Brownian dynamics side is proportional to the net number of particles which have flowed between the regions in the preceding time-step. The flux from the remainder of the PDE subdomain is calculated according to the usual centred finite-difference approximation of the diffusion equation.

The first method is then used by \citet{gorba2004bds} to investigate the behaviour of cytochrome c molecules which move in the presence of a charged membrane. Two kinds of external force are considered (electrostatic interaction and van der Waals forces) between pairs of cytochrome c molecules and between cytochrome c molecules and the charged membrane. The system is modelled as follows. The region of interest (see Figure \ref{figure:geyer_and_gorba} \subref{fig:gorba2004bds}) is a cuboid-shape box, with equal width and length. On each side of of the box, reflective boundary conditions are implemented, whilst the base of the box has a repelling boundary condition due to the repulsion caused by van der Waals forces between the membrane and the molecules. At a prescribed height there is an interface, below which particles evolve according to a Langevin equation, and above which is a fixed-density reservoir of particles. All simulations using this method are initialised with no particles in the Brownian subdomain, with particles entering solely via the reservoir.

The authors compare the results using their hybrid coupling algorithm with previous simulation results, which assume a fixed number of particles with a zero-flux boundary condition replacing the reservoir at the top of the box. They show that the shape of concentration profiles as a function of distance from the membrane generated by the two methods agree.

% \textcolor{red}{***Not sure that this paragraph is needed. I think it is a bit odd to be talking so much about a specific application, but perhaps it is inkeeping with the calcium example? We can chat about it. >> I agree, perhaps remove? << *** }

% \textcolor{red}{The ***What is the insertion method? i.e. the initialisation of Brownian particles from PDE mass?*** insertion method \citep{geyer2004ibd} is then tested using different parameter values to check sensitivity. Firstly, it is shown that the size of the membrane affects the concentration of particles bound to it, but the height of the box (i.e. distance perpendicular to the membrane) seems to make no difference. This is due to the reflective boundary conditions, which cause a build up of particles close to the walls. The authors use the effective box size to correct for this boundary effect. The effective box-size is the region of the box in which the bound concentration remains approximately constant in both directions parallel to the membrane. Finally, the effect of placing point charges within the membrane is investigated. By adding grids of charges of varying sizes to the membrane the authors show that the bound concentrations generally increase as the number of additional point charges increases.}

% (I) alexander2002ars
% (I) alexander2005ars

\begin{figure}[h!]
\centering
\includegraphics[width=0.8\textwidth]{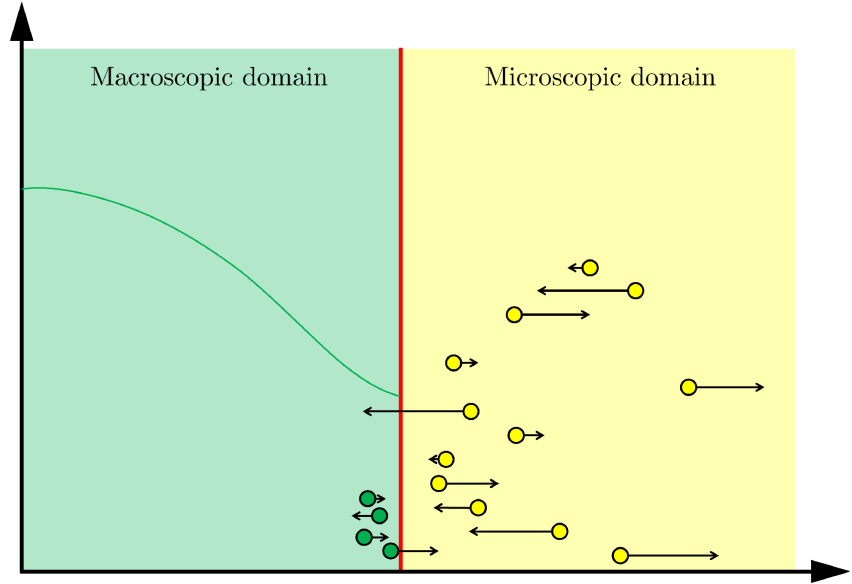}
\caption{Schematic for the method by \citet{alexander2002ars}. The green line and yellow dots represent the same phenomena as in Figure \ref{fig:ARM_Schematic}. The green dots residing within the PDE subdomain are particles initialised at the beginning of a time-step (corresponding to the numbers of particles within the corresponding region obtained by direct integration of the PDE solution). Black arrows show the directions and magnitudes of next movement of all particles. The discretisation on the lower axis is the PDE mesh over the entire domain.}
\label{fig:alexander2002ars}
\end{figure}

In contrast to the previous works presented here, \citet{alexander2002ars} introduce a hybrid method to couple an \textit{SPDE} (as well as a similar algorithm for a PDE) to Brownian dynamics (see Figure \ref{fig:alexander2002ars}). Separating the continuum and individual-based subdomains is an interface, over which particle fluxes are matched to ensure that particle movement is correctly calculated between the two descriptions. The continuum subdomain is divided into a mesh, upon which the solution to the SPDE/PDE is calculated numerically. In the particle-based subdomain, particles move according to the standard off-lattice Brownian motion SDE. The hybrid algorithm progresses in discrete time with both subdomains using the same time-step. 

In order to hybridise the two methods, at the beginning of each time-step, an integer number of particles are uniformly initialised within the SPDE/PDE voxel closest to the interface, referred to as the ``handshaking'' region. The number of particles initialised is the closest integer to the value of the SPDE/PDE solution at the handshaking mesh point at the beginning of the time-step. All particles (both in the handshaking region and elsewhere) are then evolved according to the standard Brownian motion equation. The number of particles crossing the interface gives the flux into the handshaking mesh point which is stored and later implemented when the PDE/SPDE values are updated. Any particles which do not reside in the Brownian subdomain following the position update step are removed from the simulation. All other SPDE/PDE fluxes are calculated using the discretised version of the SPDE/PDE equation and the values of the mesh points are consequently updated.

In a later paper, the same authors also consider correlated systems \citep{alexander2005ars}. They develop a hybrid algorithm for the train model which describes the transport of material in a viscous gas. This model is chosen due to its relative simplicity and the readily derived continuum (SPDE/PDE) counterparts which are straightforward to solve numerically. The train model can be summarised as follows: several trains run parallel to one another at different speeds with varying numbers of passengers. Passengers jump, with exponentially distributed waiting times, between neighbouring trains, changing the momentum of the participating trains. At each end of the array of trains are ``platforms'' which move at a fixed velocity and contain a reservoir of passengers. 

The authors couple a discretised version of the SPDE/PDE representation of the train model to the discrete individual-based description. Both the discretised SPDE/PDE and the train model are simulated with the same grid spacing. Separating the two subdomains is an interface. The hybrid algorithm uses flux-matching for both the velocity and the momentum over the interface, whilst also maintaining the long-range spatial correlations in the velocity caused by stochastic fluctuations. The algorithm employed is analogous to the one that is presented in \citet{alexander2002ars}. At the beginning of a continuum time-step the first voxel in the continuum part of the domain (called the ``handshaking'' region) is filled with particles.  The number of particles initialised is the nearest integer value to the SPDE/PDE solution in this voxel. Each of these particles is also assigned a velocity which corresponds to the velocity of the continuum model at that point. The individual-based particles are then evolved and the fluxes of velocity and momentum over the interface are calculated. These values are then utilised within the continuum solver in place of the the fluxes over the interface.

% plapp2000mrw (here or in other?)
Finally, \citet{plapp2000mrw} introduce a hybrid method for simulating interfacial patterns, with specific application to dendritic crystal growth. In the inner-region, which includes the area in which the crystal is growing and a buffer layer of liquid adjacent the interface, a discretised version of the diffusion equation is solved and the position of the crystal interface is updated using a deterministic phase-field approach. This update method is coupled to particles evolving according to off-lattice Brownian motion. The time-step at which the positions of particles are updated increases the further away the particles are from the interface. At the edge of the inner-region between the crystal surface and the outer region is a ``buffer-region'' of undercooled liquid. This buffer-region acts to damp the stochastic variation of the outer-region to negligible levels at the crystal surface. Adjacent to the interface between in the inner and outer regions are ``conversion cells'' which facilitate the conversion of Brownian walkers into PDE density and vice versa, via the implementation of boundary conditions on each of the models.  A Dirichlet boundary condition for the PDE is determined by the number of Brownian particles residing in each of the conversion cells. In the other direction, the heat flux over the boundary is collected in a reservoir. If the value of the reservoir exceeds a threshold, $H$, a new particle is added to the cell. If it drops below $-H$, then a particle is absorbed and consequently removed from the corresponding conversion cell.

\section{Other hybrid methods} \label{sect:Other}

Within this section, we investigate some other hybrid methods that do not fall within any of the above three categories. The section will encompass microscopic-to-molecular dynamics spatially-coupled methods. These hybrid methods are typically designed to represent hydrodynamical systems, adaptive mesh and algorithm refinement and quasicontinuum methods. We will also investigate another class of hybrid methods, which we shall call ``species splitting'', where different species are simulated using different representations. 

%% Micro-molecular models
\subsection{Micro-molecular methods}

In this subsection, we present a paper which introduces hybrid methods for coupling a molecular dynamics model to a corresponding Brownian motion model for the movement of a large particle in a surrounding `molecular' medium. 

%(I) erban2014fmd

\citet{erban2014fmd} introduces one such spatial hybrid method in one and three dimensions. The author motivates the use of such a method by considering a large focal protein molecule which is being moved by interactions with the smaller water molecules that surround it. The protein molecule is modelled as a hard sphere with a larger radius and mass than the water molecules. The motion of the molecules in this molecular dynamics model are fully deterministic once they have been randomly initialised, with changes in velocity caused by momentum exchange. If the protein molecule were to be modelled using Brownian dynamics or the Langevin equation (respectively), the interactions between it and the surrounding water molecules could be encapsulated implicitly through the random changes in position or velocity (respectively) of the protein. \citet{erban2014fmd} demonstrates the equivalence between the motion of the protein molecule in the molecular dynamics simulation to the motion specified by the corresponding Langevin or Brownian dynamics equations in certain limits. This equivalence engenders the possibility of a hybrid method.
 
In both the one- and three-dimensional hybrid methods, the domain is split into two subdomains: one in which water molecules are explicitly simulated and the other in which the water molecules are modelled implicitly and the protein moves according to the appropriate Langevin equation.
The first coupling algorithm introduced is for a one-dimensional domain, in which water molecules are initialised across a subset of the real line according to a spatial Poisson point process with a specific density, while velocities are normally distributed with zero mean and variance which incorporates the diffusion coefficient, the ratio between the large and small particles' masses and a friction coefficient. Collisions between water molecules and proteins are elastic and subject to conservation of momentum. Any water molecules which leave the molecular dynamics subdomain are removed from the system. Molecular dynamics particles can also be created towards the edges of the subdomain, and are initialised using a normalised complementary error function. This maintains the density of water molecules in the molecular dynamics heat bath. The three-dimensional algorithm is similar. The algorithms are time-driven, that is the system is evolved by implementing exchange of momentum through collisions, updating positions and the addition and removal of heat bath molecules at each fixed time-step. There is a constraint on the size of the time-step to ensure that at most one macro particle  enters the subdomain in each time-step. A similar coupling is presented in \citet{erban2016caa}.

\begin{figure}[h!]
\centering
\includegraphics[width=0.8\textwidth]{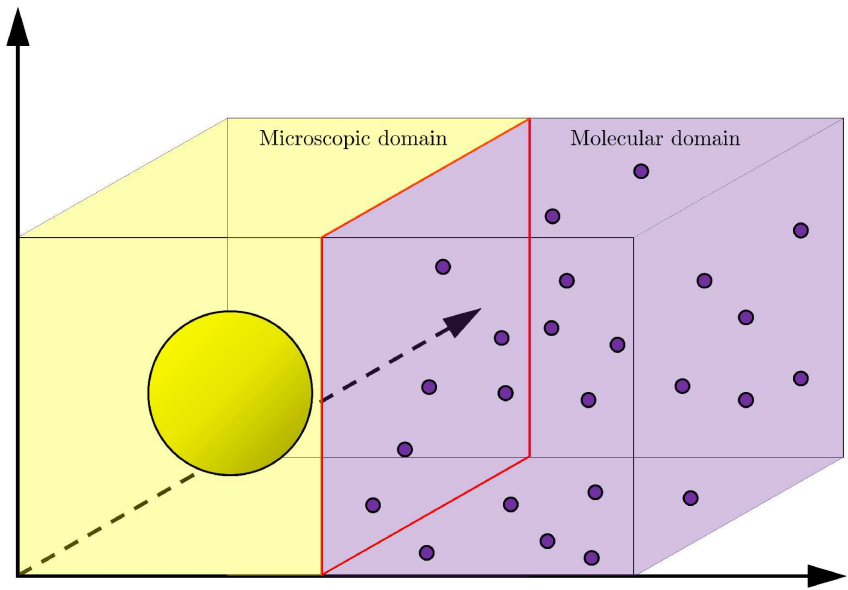}
\caption{A schematic for the method presented by \citet{erban2014fmd}. The large yellow circle is an individual particle (protein molecule) with mass, volume and velocity. The small purple dots represent the molecular dynamics particles (air/water molecules) and also have a mass, volume and velocity.}
\label{fig:erban2014fmd}
\end{figure}

% erban2016caa

%% Hydrodynamics. However, some of the papers in the previous three sections are hydrodynamics, may need to consider placing these in the main three section.
% donev2010hpc
% flekkoy1999mdd
% flekkoy2000hmc
% wagner2004hcf
% wagner2002cmd
% delgado2005hmc x3
% hadjiconstantinou1999cac
% hadjiconstantinou1999hac
% mohamed2010rdh
% koumoustakos2005mfs
% oconnell1995mdc
% wijesinghe2004dha
% williams2008arf

\subsection{Hydrodynamics}

Whilst most of the examples that have been presented in Sections \ref{sect:MacroMeso}-\ref{sect:MacroMicro} are designed to represent reaction-diffusion systems (with noted exceptions), these are not the only systems in which spatial hybrid methods have been employed. In this subsection we review spatial hybrid methods and their uses in modelling hydrodynamics in an efficient and accurate manner.

The most common type of spatially-coupled hybrid method employed within hydrodynamics is macro-micro couplings. \citet{donev2010hpc} couple the stochastic hydrodynamics model given by the Landau-Lifshitz Navier-Stokes (LLNS) equations, to a corresponding direct simulation Monte-Carlo representation. The LLNS equations include hydrodynamic fluctuations, and, as such, are SPDEs. They are simulated using a fixed-time, three-stage Runge-Kutta integration scheme (a finite volume method) although the authors note that other other finite-volume explicit schemes can be substituted. Within the particle subdomain, the hydrodynamics are simulated using a fixed-time stochastic momentum exchange method which preserves the essential hydrodynamic properties of molecular dynamics. The time-scale of the micro solver is smaller than that of the macro solver, so that multiple particle updates occur for every continuum update. This is in contrast to PDE-assisted Brownian dynamics \citep{franz2012mrd} for reaction-diffusion systems which does the opposite.

Within the continuum subdomain, the only quantities that need to be considered are the conserved variables of mass, momentum and energy within each continuum cell, as well as the continuum normal flux between any two neighbouring macroscopic cells. Within the particle subdomain, inter-atomic forces are simulated by stochastic collisions, so that any particles within a given distance have a probability of colliding. Separating the two subdomains is an (adaptive) interface. The coupling algorithm ensures that both the fluxes and the states (density, momentum and energy) at the interface are continuous by introducing a state-flux coupling methodology; the macroscopic LLNS equations act as a source of particles into the microscopic subdomain at the interface, and the particles impose a flux boundary condition on the continuum. To impose the state boundary condition from the continuum subdomain onto the particle subdomain, a reservoir of temporary particles (in a small region within macro cells adjacent to the interface) are initialised (every micro time-step) with some velocity and temperature according to a Maxwell-Boltzmann or Chapman-Enskog distribution chosen to match the velocity and temperature of the associated macro cell (reminiscent of the method of \citet{alexander2002ars} for modelling diffusion). The number of these particles is chosen to match the continuum density in the associated macro cell. The particle flux over the interface is calculated and stored every micro time-step and imposed on the continuum solver at the end of every macro time-step.

There are other methods which also utilise an interface in order to couple two subdomains. \citet{flekkoy1999mdd} couple the mesoscopic dissipative particle dynamics to the derived Langevin equation in order to simulate the movement of large colloid molecules. \citet{oconnell1995mdc} also utilise an interface in order to create a generic algorithm for simulating a macroscopic and microscopic representation of a fluid system. The authors couple by averaging the velocities of the individual particles close to the interface, providing a boundary condition for the corresponding continuum model.

Overlap regions have also been employed in the hydrodynamics literature. \citet{flekkoy2000hmc} couple a macroscopic PDE to a microscopic method in which particles interact according to Lennard-Jones potentials \citep{allen2017csl}. Separating the two subdomains is an overlap region in which both the particle and continuum descriptions are valid. The conservation of mass and momentum between the two regions is handled explicitly using flux exchange, which means that the coupling scheme adheres to the relevant conservation laws.

Within the continuum description, the mass and momentum fluxes are represented using finite-differences across each continuum node. These finite-difference approximations are used to advance the continuum equations in time. The boundary conditions derived from the particle region are implemented on the continuum representation by replacing the fluxes at the end of the continuum subdomain with the mean mass and momentum fluxes of particles around the boundary, averaged over a continuum time-step. To implement the fluxes of mass and momentum from the macroscopic to microscopic subdomain, a number of particles per unit time (determined in order to conserve mass flux) are placed into a region close to the boundary of the particle subdomain. Additionally, the velocities of the particles are chosen to conserve the flux of momentum. The authors note that there is an asymmetry relating to the fluctuations using their method; the continuum subdomain effectively acts to damp fluctuations in the particle subdomain  meaning, for example, that fluctuations in particle numbers will be diminished in comparison to predictions from statistical mechanics (reminiscent of the damping of the Brownian dynamics by the PDE observed by \citet{franz2012mrd}).

A second coupling, presented by \label{page:flekkoy_name} \citet{wagner2004hcf} extends previous works \citep{wagner2002cmd,flekkoy2000hmc}, in which fluxes for momentum and mass were preserved between the two subdomains, to the situation in which energy flux is also conserved. The authors also investigate the limitations of this hybrid representation when  simulating both homogeneous and gradient flow. 

The continuum equations are discretised using a centred finite-difference scheme on a regular mesh. Separating the continuum and particle subdomains is an overlap region which allows for the conservation of flux between the two descriptions.
To calculate the continuum flux for the penultimate node within the overlap region (which corresponds to the boundary of the particle subdomain), a similar method to the one employed by \citet{flekkoy2001cpf} is used. One of the terms in the centred finite-difference approximation is replaced by the corresponding value from the particle subdomain at the particle end of the overlap region. These fluxes (for mass momentum and energy) are then arithmetically averaged with the corresponding mean fluxes of the particles that occupy positions within the final voxel of the overlap region. These mean fluxes are then used to implement Neumann boundary conditions on the final node of the continuum representation.
The same averaged fluxes are implemented on the particle subdomain by adding/removing particles to/from the microscopic description in a region corresponding to the penultimate node of the continuum discretisation. To ensure that both momentum and energy are conserved, velocities and accelerations of particles in the overlap region are altered accordingly.

Several other papers have adopted the use of an overlap region. \citet{wagner2002cmd} use mutual flux exchange in order to couple their finite-difference representation of a PDE for fluid flow to the corresponding microscopic dynamics. The authors measure the fluxes for mass, momentum and energy in order to ensure conservation. \citet{delgado2005hmc} and \citet{delgado2005fcp} present two further papers which couple using flux conservation. These methods use flux exchange from the continuum to particle density in order to modify the microscopic description, while fluxes in the opposite direction supply boundary conditions for the continuum representation.

\citet{delgado2009cac} present a hybrid method with three spatial scales - coupling the macroscopic to the mesoscopic to the microscopic scales, with an application to liquid water. The authors use two different schemes in order to complete the coupling. To couple between the macro and microscales, the HybridMD scheme is used \citep{defabritis2006mml} and to couple the microscale to the mesoscale, the adaptive resolution scheme (AdResS) is employed \citep{praprotnik2005arm}.

There are many other papers which have addressed hybrid methods for hydrodynamics. We direct the interested reader to the reviews of \citet{koumoutsakos2005mfs} and \citet{mohamed2010rdh} and the PhD thesis of \citet{hadjiconstantinou1999hac} for further details.

%% Well-mixed. Maybe for in the introduction to simply reference
% bobashev2007hem
% duncan2016hfs
% hellander2007hmc
% burrage2004msa
% bentele2004gsh
% hepp2014ahs
% Kiehl2004hsc
% salis2005ahs

%% Algorithm mesh refinement
% garcia1999ama
% williams2008arf

%% Quasicontinuum methods
% shenoy1999afe
% tadmor1996qad

\subsection{Adaptive mesh and algorithm refinement}

Adaptive mesh refinement (AMR) is a method for evaluating PDE solutions on inhomogeneous domains, in which coarse cells are recursively refined in both time and space in regions of high sensitivity \citep{berger1989lam}. Adaptive mesh and algorithm refinement (AMAR) extends the idea of AMR. The difference between AMR and AMAR is that when the predefined highest spatial resolution has been reached, AMAR switches to using a discrete method for simulating the underlying phenomena. The coupling between the coarse PDE and the fine discrete method is completed using a buffer region residing within the PDE region close to the interface between the two subdomains. Particles are created within this region at the beginning of the fixed PDE update time-step with the appropriate physical quantities such as mass, momentum and energy, and are then allowed to flow forwards in time. This provides boundary conditions for the two systems. \citet{garcia1999ama} and \citet{williams2008arf} use AMAR in order to accurately model hydrodynamic flow.

\subsection{Quasicontinuum methods}

Quasicontinuum (QC) methods  combine continuum and atomistic representations for modelling crystalline structures, and were first introduced by \citet{tadmor1996qad}. \citet{shenoy1999afe} propose a hybrid method for coupling the atomistic-scale dynamics of solid deformation to a corresponding continuum description. The quasicontinuum method exploits the kinematic constraints inherent to the atomistic lattice, reducing the large number of degrees of freedom by employing the finite-element method in order to simplify the minimisation of the potential energy associated with the system under a deformation.
The system of interest is typically made up of a huge number of atoms, and consequently has an extremely large number degrees of freedom. It is therefore computationally difficult to calculate any quantity of interest. To reduce the number of degrees of freedom, a subset of the atoms are chosen to be \textit{representative atoms}. Each representative atom is a proxy for a number of neighbouring atoms, reducing the number of degrees of freedom. Close to the deformation, where each atom experiences a different local environment, atoms are represented individually. In these regions, an atomistic, non-linear approach to calculating the energy is required. Further from the deformation, where non-linear effects are negligible and each representative atom is a proxy for some of its neighbours, linear elasticity theory is used. This allows for the faster calculation of the energy landscape in large regions of the spatial domain without the loss of accuracy in the regions in which a more detailed representation is required. The condition which specifies the homogeneity, or otherwise, of a local region is determined by calculating the right stretch tensor of the deformation. If the maximum difference of the eigenvalues over any pair of atoms within a given distance is less than a pre-determined threshold, it is treated as a near-homogeneous environment. This ensures that the algorithm adaptively chooses which regions are to be treated as homogeneous. However, the algorithm does create additional forces, referred to as ``ghost forces'', due to the hybridisation. These are corrected for by applying correction forces within the energy minimisation calculation.

%% Equivalence Frameworks
% doering2003ips
% espanol1995hdp
% hoogerbrugge1992smh ???
% bernstein2005smr ??? AMR?
% engblom2009ssr  ??? Hybrid?
% bruna2012eve
% chen2012fbd (spatial -> non-spatial)
% erban2009smr
% koumoustakos2005mfs
% franz2013twh
% isaacson2009rdm
% isaacson2013crd

%% Other hybrids (Different species consistently different descriptions)
% anderson2005hmm
% anderson1998cdm
% dorman2002mso
% franz2011hmi <<<<<<<<<<<<<<< Revisit
% franz2013twh
% gerlee2007ehc
% jackson2006ldp
% jeon2010olh
% taylor2016cve (Volume exclusion -> partial volume exclusion)
% wylie2006hds
% jeschke2008mrs

\subsection{Other hybrids}

This section contains several hybrid methods that do not fall into the spatially-coupled reaction-diffusion, or hydrodynamics categories. They are designed to model a wealth of different mathematical, biological and physical problems and employ a variety of hybridisation techniques.

\citet{jeschke2008mrs} introduce a hybrid method for the simulation of  macromolecular crowding. They combine the mesoscopic next subvolume method (NSM) \citep{elf2004ssb} for the efficient simulation of compartment-based reaction-diffusion systems with an off-lattice representation of large crowding particles (crowders). The crowders are spherical and evolve according to an individual-based method which assumes random movements of particles over fixed time-intervals. All other particles are updated using the NSM on a square lattice. 

Crowders occupy a certain volume. As they move, the volume that is available for the compartment-based particles and their interactions changes. Any compartments which intersect a crowder are sub-divided, using an octree refinement algorithm, until a pre-defined number of sub-divisions have been completed. The volume of the compartment that is occupied by the crowder is then approximated as the number of sub-octants that intersect the it. The crowders and compartment-based particles can interact with one another. For example, the location of overlapping crowders will influence the neighbouring compartments into which compartment-based particles are able to diffuse. Diffusion occurs at the usual diffusive rate, but scaled down by the proportion of the boundary between the current compartment and the neighbouring compartments that is occupied by crowders. Particles can also bind to the crowders, meaning that they are removed from the NSM reactions list and move about with the crowder. When the crowders move, they ``push'' the compartment-based particles into the unoccupied region of their current compartment, or into neighbouring compartments if the crowder completely fills their current compartment. All movements, reactions and steric interactions are controlled by the ``coordinator component'' which keeps track of all putative next event times, schedules the next reaction and updates the two systems.

There are many spatially-extended hybrid methods in which some species are represented using continuum models throughout the domain and others using discrete models in the same domain. These methods are popular when representing species which are inherently different in copy number throughout the domain. For example, small numbers of chemotaxing bacterial cells might be represented using an individual-based model, whereas the chemical signal to which they respond might be represented as a continuum. Since these models are not the primary focus of this review (rather we focus on models in which the same species is represented variably throughout the domain) we will give only a brief mention to some of these hybrid methods.

Cancerous tumour behaviour has frequently been represented using such hybrid methods. \citet{anderson1998cdm} model angiogenesis -- the directed growth of blood vessels towards the tumour. In order to do so they couple the macroscopic system of PDEs governing the growth of a tumour to a discrete model of blood vessel formation on a lattice. The discrete model is used in order to investigate how individual cells branch and undergo  anastomosis and mitosis close to the tips of blood vessels which have sprouted. The authors also use a similar method to model the invasion of healthy tissue by a solid tumour \citep{anderson2005hmm}. Other examples of tumour growth hybrid methods include the use of cellular automata \citep{dormann2002mso,gerlee2007ehc} and a method which models the environment as a continuum, while the tumour cells themselves are discrete and react the environment \citep{jeon2010olh}. A similar idea has also been employed by \citet{franz2013twh}, in which bacteria respond to a chemotactic signal. The signal is modelled by a continuum PDE, which the bacteria, modelled as individuals, can adapt and respond to.

%% Introduction
% drawert2010dfs
% drawert2010urd
% hattne2005srd
% pahle2009bss (contains all software references, URDME PYURDME)

\section{Discussion and outlook} \label{sect:Discussion}

% To include:
%	Summarise the importance of the methods
%	Lots of them - what can they deal with
%	Other types of hybridisation (non-spatial) - beyond scope of the review
%	What is being done now? (Erban micro-molecular, variances, ARM)
Within this review, we have explored the rich and diverse field of spatial hybrid methods, and illustrated how they can be utilised in order to probe previously intractable problems in the biological and physical sciences. Biological and physical phenomena exists at a variety of temporal, spatial and population scales \citep{sherratt2005avs,volpert2009rdw,mort2016rdm,khan2011scd, dobramysl2015pbd}.
\label{page:conc_scales_con}
Take, for example, the formation of calcium puffs at the endoplasmic reticulum \citep{dobramysl2015pbd}. Just before a calcium ion channel opens, the number of calcium ions is small. However, once the channel opens, the number of particles becomes orders of magnitude larger. Further away from the channels, particle numbers remain relatively small until diffusion disperses them. Even for a single phenomenon, populations can vary over orders of magnitude making traditional modelling approaches difficult. Novel modelling methods which span these scales in a computationally efficient manner may provide insights into these phenomena. This is precisely the purpose of many of the hybrid methods reviewed in this paper -- they permit the representation of multiple scales within a system, allowing for efficient and accurate simulation. This review has focussed mostly on spatially-coupled hybrid methods for reaction-diffusion systems that allow space to be partitioned into subdomains in which different modelling paradigms are employed.

We covered couplings that broach four different spatial scales -- the macro, meso and microscales, together with molecular dynamics. We have provided detailed summaries of illustrative examples for macroscopic-to-mesoscopic (PCM by \citet{yates2015pcm}), mesoscopic-to-microscopic (GCM by \citet{flegg2015cmc}) and macroscopic-to-microscopic (ARM by \citet{smith2017arm}) couplings, together with pseudocode for their implementation and demonstrations of worked examples, in order to facilitate the use of such hybrid methods. In addition, in the electronic supplementary material for this paper we provide working MATLAB code for each of the three methods. Schematics and descriptions of various other methods provide an extensive yet non-exhaustive list of possible hybrid methods, which should be chosen depending on the application at hand, and the type of coupling desired.

Whilst not the focus of this review, there are other hybrid methods in which space is not modelled explicitly. Several hybrid methods concern the simulation of well-mixed chemical systems \citep{duncan2016hfs,hellander2007hmc, burrage2004msa, bentele2004gsh, salis2005ahs} while epidemiology \citep{bobashev2007hem} and stochastic reaction networks \citep{hepp2014ahs} have also been investigated. We have also described several spatially-extended methods which used different types of hybridisation within section \ref{sect:Other}.

This review contains a summary of the current state of spatial hybrid methods. We now look to the future and directions in which the area will progress. Whilst much work has been completed within the field, there are still issues which are common to many of the methods. Chief amongst these is variation in hybrid methods that involve deterministic PDEs compared to the full solution simulated using a stochastic approach. Typically the deterministic nature of the continuum model results in damping of the variation in the stochastic subdomain in comparison to that of the fully stochastic method. Some authors have fixed this problem by incorporating an overlap region instead of an interface \citep{harrison2016hac, franz2012mrd, flekkoy2001cpf}. Within the overlap region, mass is simultaneously modelled using both representations. A second method for resolving the variance is to replace the PDE with an appropriate SPDE, a macroscopic model for which stochasticity is inherently incorporated. Provided the stochasticity is chosen in a consistent manner (consistent with the fully stochastic method), hybrid methods have been postulated for which the variance in the individual subdomain has been shown to match that of the fully stochastic model \citep{alexander2002ars}. 

As mentioned in Section \ref{sect:Other}, recently there has been work to couple microscopic descriptions to molecular dynamics. \citet{erban2014fmd, erban2016caa} has pioneered work in this area, providing methods which do just this. This type of method can be utilised in order to simulate biological phenomena at the molecular level, which even microscale Brownian motion may be unable to accurately capture.

There is a relative abundance of spatial hybrid methods (attested to by this review). Although we have presented a small number of papers which employ these methods in real physical and biological problems, there still remain very few practical applications of such methods. Whether this is due to the complexity of the hybrid methods in comparison to their single model counterparts or to the low profile of such methods, the challenge remains for the developers of such hybrid algorithms to realise the potential impact of their methods by applying them to real problems. We hope that this review has served the purpose of increasing the profile of hybrid methods, whilst simultaneously making them more accessible to the user.

\section*{Author contributions}
CS and CY contributed equally to the production of this manuscript. CS performed the simulations and created the figures.

\section*{Acknowledgements}
We would like to thank the anonymous reviewers for their constructive comments on the manuscript.

\section*{Data accessibility}
The accompanying code has been uploaded as part of the supplementary material.

\section*{Funding}
CS is supported by a scholarship from the EPSRC Centre for Doctoral Training in Statistical Applied Mathematics at Bath (SAMBa), under the project EP/L015684/1.

\section*{Competing interests}
We have no competing interests.

\bibliographystyle{plainnat}
\bibliography{hybrid_review}

\end{document}